\documentclass[aps,prb,twocolumn,superscriptaddress]{revtex4-1}
\usepackage{amsmath}
\usepackage{amssymb}
\usepackage{graphicx}
\graphicspath{{figures/}}
\usepackage{verbatim}
\usepackage{amsfonts}
\usepackage{dcolumn}
\usepackage{bm}
\usepackage{color}
\usepackage{multirow}
\usepackage{tabularx}
\usepackage{pifont}
\usepackage[colorlinks,citecolor=blue]{hyperref}
\usepackage{slashed}
\usepackage{ulem}

\usepackage{mathrsfs}

\newcommand{\beq}{\begin{eqnarray} }
\newcommand{\eeq}{\end{eqnarray} }
\newcommand{\Beq}{\begin{eqnarray*} }
\newcommand{\Eeq}{\end{eqnarray*} }
\newcommand{\Bmat}{\left(\begin{matrix}}
\newcommand{\Emat}{\end{matrix}\right)}
\newcommand{\up}{\uparrow}
\newcommand{\dn}{\downarrow}

\begin{document}

\title{Magnon condensation in dimerized antiferromagnets with spin-orbit coupling}

\author{Qi-Rong Zhao}
\affiliation{Department of Physics, Renmin University of China, Beijing 100872, China}

\author{Meng-Jie Sun}
\affiliation{Department of Physics, Renmin University of China, Beijing 100872, China}

\author{Zheng-Xin Liu}
\email{liuzxphys@ruc.edu.cn}
\affiliation{Department of Physics, Renmin University of China, Beijing 100872, China}
\affiliation{Tsung-Dao Lee Institute \& School of Physics and Astronomy, Shanghai Jiao Tong University, Shanghai 200240, China}

\author{Jiucai Wang}
\email{wangjiucai@mail.tsinghua.edu.cn}
\affiliation{Institute for Advanced Study, Tsinghua University, Beijing 100084, China}
\affiliation{Department of Physics, Renmin University of China, Beijing 100872, China}

\date{\today}

\begin{abstract}
Bose-Einstein condensation (BEC) of triplet excitations triggered by a magnetic field, sometimes called magnon BEC, in dimerized antiferromagnets gives rise to a long-range antiferromagnetic order in the plane perpendicular to the applied magnetic field. To explore the effects of spin-orbit coupling on magnon condensation, we study a spin model on a distorted honeycomb lattice with dimerized Heisenberg exchange ($J$ terms) and uniform off-diagonal exchange ($\Gamma'$ terms) interactions. Via variational Monte Carlo calculations and spin-wave theory, we find that an out-of-plane magnetic field can induce different types of long-range magnetic orders, no matter if the ground state is a non-magnetic dimerized state or an  ordered N\'{e}el state. Furthermore, the critical properties of field-driven phase transitions 
in the presence of spin-orbit couplings, as illustrated from spin-wave spectrum and interpreted by effective field theory, can be different from the conventional magnon BEC.
Our study is helpful to understand the rich phases of spin-orbit coupled antiferromagnets induced by magnetic fields.

\end{abstract}

\maketitle

\section{Introduction}

\begin{figure}[b]
\includegraphics[width=8cm]{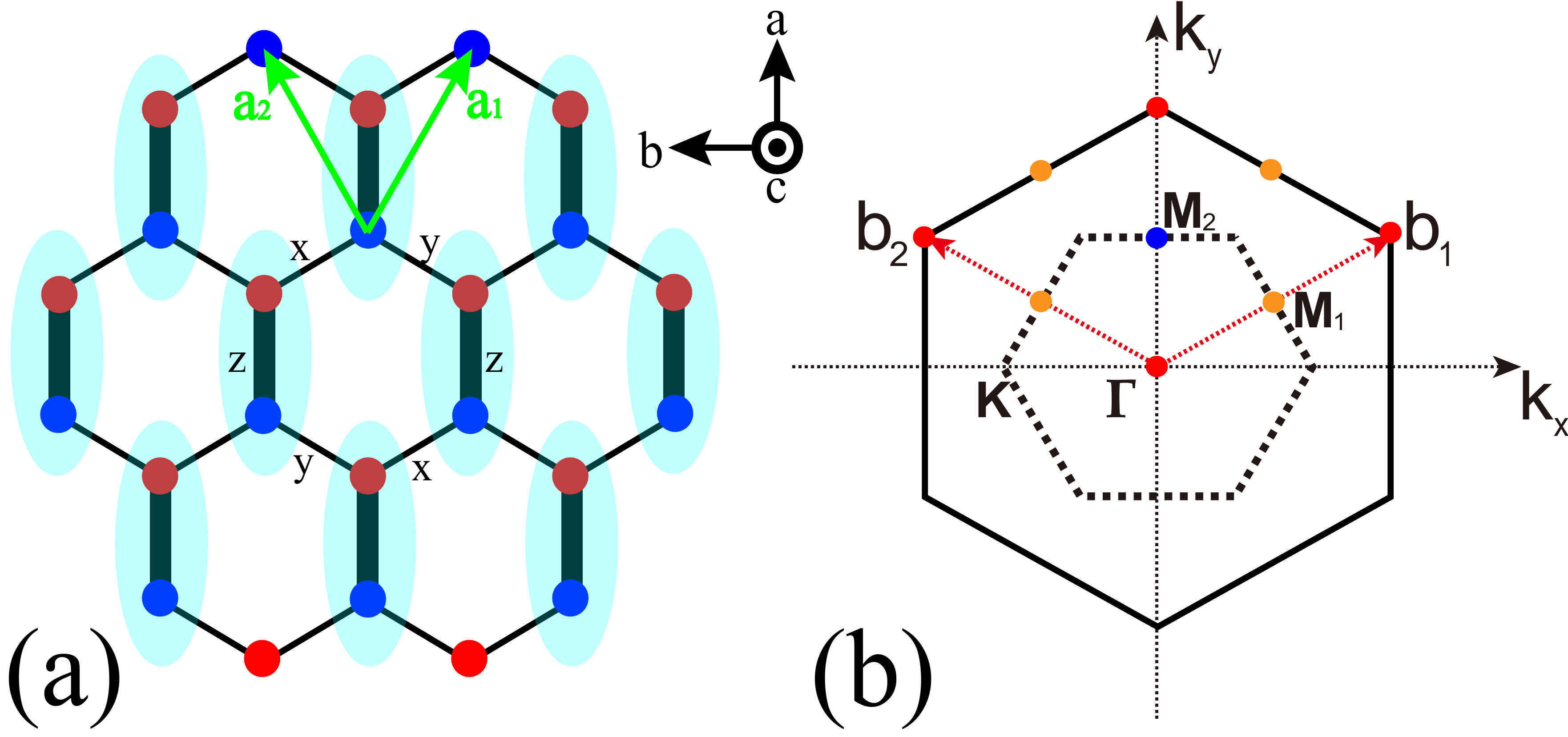}
\caption{
(a) The x, y, and z bonds in $J$-$\Gamma'$ model are shown and the thickness of bonds shows the strength of Heisenberg ($J$) interactions.
The elliptic shadings stand for singlets formed by dimerized spins.
The crystallographic a (in-plane), b (in-plane), and c (out-of-plane) directions are indicated, whose directions are given by [1$\overline{1}$0], [11$\overline{2}$], and [111], respectively. (b) The corresponding Brillouin zone (BZ) of a honeycomb lattice spanned by unit vectors $\vec{a}_1$ and $\vec{a}_2$.
}\label{fig:hd}
\end{figure}

Quantum magnetism, which involves interacting spins and orbitals in magnetic materials, is a good platform for investigating many-body physics. The quantum dimer magnet (QDM), a nonmagnetic ground state constituted by distinct singlet-dimers, is one of the simplest examples of quantum magnets having no classical correspondence. An example of QDMs is the antiferromagnetic (AFM) Heisenberg model on the distorted honeycomb lattice, where the thicker bonds (e.g. the vertical bonds, see Fig.\ref{fig:hd}(a) for illustration) form disconnected dimers. Since the interactions on the thinner bonds are relatively weaker, the ground state can be approximately considered as a direct product of singlets formed by the dimers. The excitation on each singlet dimer forms a triplet, which is bosonic and sometimes called a ``magnon''.

It was proposed that the magnons in QDMs can undergo BEC\cite{Giamarchi1999} just like usual Bose gas\cite{Matsubara1956}.
An external magnetic field can adjust the `chemical potential' of magnons, consequently, a magnon-BEC phase appears in the intermediate field region, which is sandwiched by the dimerized phase at the low-field side and the polarized phase (PP) at the high-field side.
The magnon BEC was experimentally verified later\cite{Nature2003, Demokritov2006, Sherman2003, Kolezhuk2004, Ruegg2007, Rosch2007, Kramer2007, Autti2012, Fang2016, Bozhko2016}. As is known that the BEC of usual bosons spontaneously breaks the particle number conserving global U(1) symmetry, similarly the BEC of magnons spontaneously breaks the $SO(2)$ spin rotational symmetry \cite{Giamarchi2008, Zapf2014}. Here the $SO(2)$ group comprises continuous spin rotation along the direction of the magnetic field.
The spontaneous breaking of $SO(2)$ symmetry indicates the appearance of long-range antiferromagnetic order in the plane perpendicular to the field direction, and the orientation of magnetic momentums is determined by the ‘global phase’ of the magnon condensate.

\begin{figure*}[t]
\includegraphics[width=18cm]{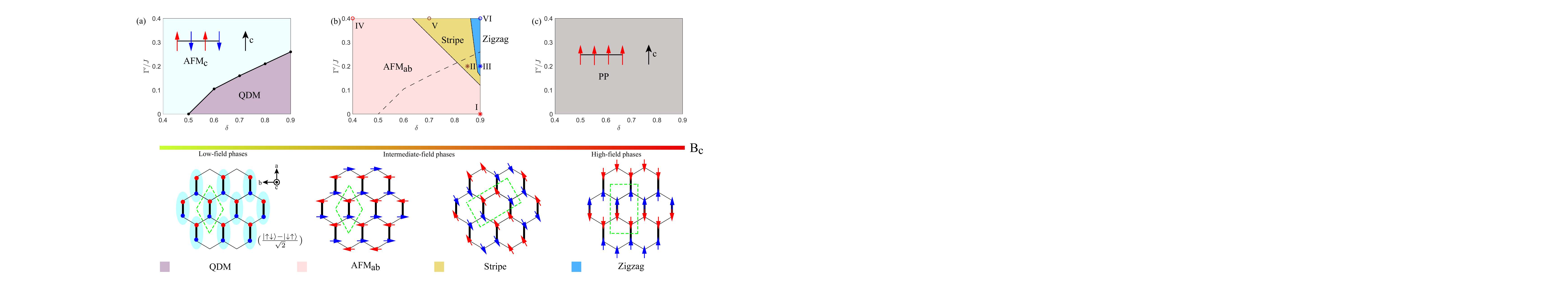}
\caption{
Schematic phase diagram of the $J$-$\Gamma'$ model on the distorted honeycomb lattice in a magnetic field along c-direction.
(a) The low-field quantum phase diagram contains two phases, the QDM phase (lower) and the AFM$_{\rm c}$ phase (upper), separated by a continuous transition (illustrated by the solid line). In the AFM$_{\rm c}$ phase, the magnetic momentum is parallel to the $c$-direction. (b) With increasing of field strength, different canted deformation of magnetic orders (including AFM$_{\rm ab}$, stripe, and zigzag) appear at the intermediate field region, accordingly the QDM (AFM$_{\rm c}$) phase splits into three different pieces labeled with I, II, III (IV, V, VI). (c) A polarized phase (PP) with magnetic momentums along $c$-direction in the high-field phase diagram.
Representative ground-state spin configurations in the corresponding phase of phase diagrams are shown below.
}\label{fig:PD01}
\end{figure*}

\begin{figure}[t]
\includegraphics[width=8cm]{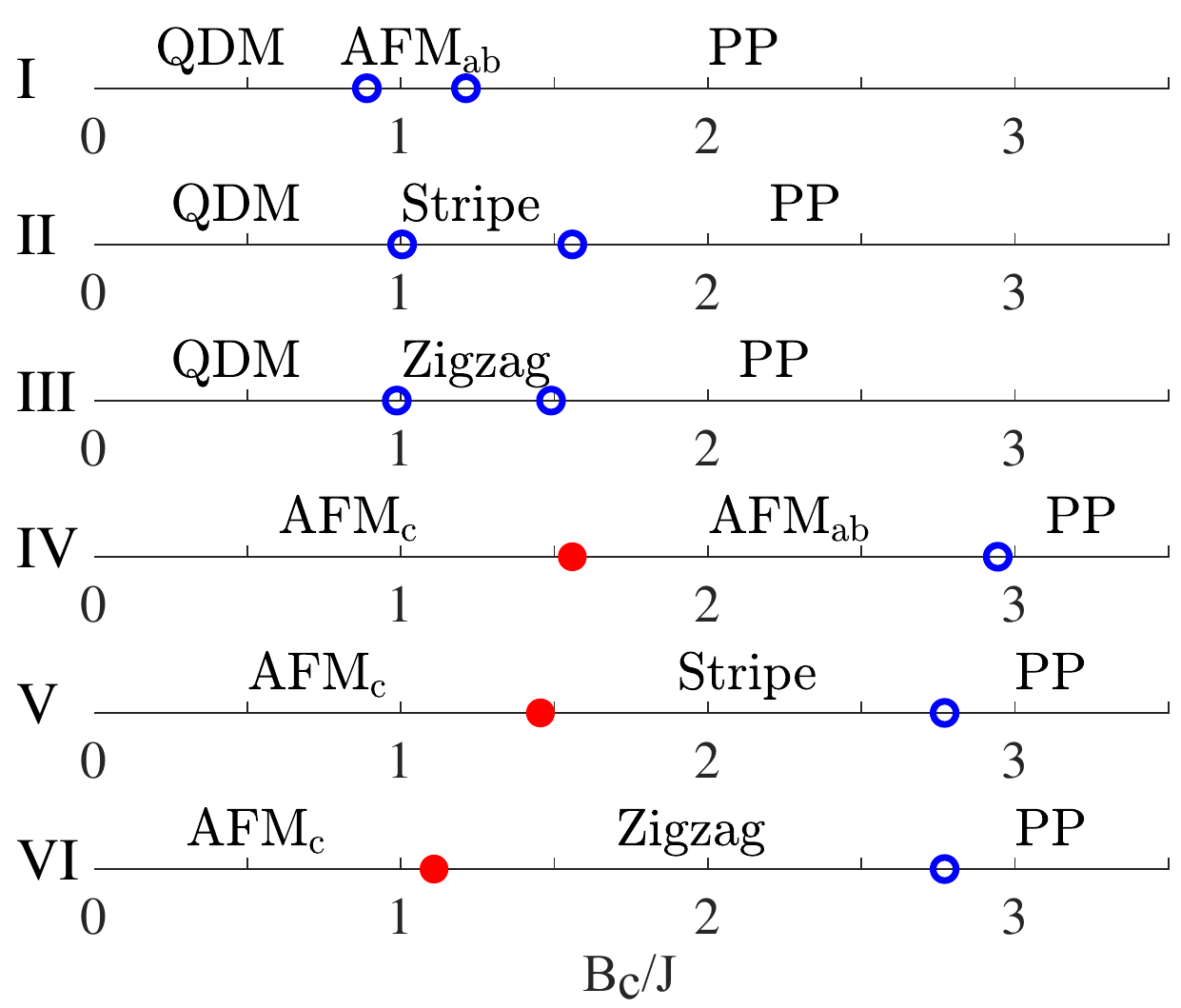}
\caption{ Phase diagrams of the $J$-$\Gamma'$ model in the magnetic field along c-direction with I: $\delta=0.9$, $\Gamma'/J=0$, II: $\delta=0.85$, $\Gamma'/J=0.2$, III: $\delta=0.9$, $\Gamma'/J=0.2$, IV: $\delta=0.4$, $\Gamma'/J=0.4$, V: $\delta=0.7$, $\Gamma'/J=0.4$, and VI: $\delta=0.9$, $\Gamma'/J=0.4$.
I$\sim$III show phase diagrams with increasing fields, corresponding to the three stellate points in region-I$\sim$III of Fig.\ref{fig:PD01}(b), respectively, while IV$\sim$VI are phase diagrams for the three circles in region-IV$\sim$VI of Fig.\ref{fig:PD01}(b), respectively.
The red solid (blue hollow) points represent first-order (continuous) phase transitions.
}\label{fig:PD02}
\end{figure}

Generally, in the real magnetic materials, the continuous $SO(2)$ symmetry explicitly breaks down into its discrete subgroups if the system contains anisotropic interactions, such as dipole-dipole interactions or spin-orbit coupling (SOC)\cite{Sirker2004}.
If the energy scale of the factors that cause the anisotropy is very weak compared to the ordering temperature, the low energy physics is approximately described by $SO(2)$ conserving theory\cite{Radu2006, Batista2007, Schmalian2008}. Therefore, it is expected that the physical picture of magnon BEC still works under an external magnetic field.
If the anisotropy of magnetic properties is strong enough, the physical consequences of the applied magnetic field will be modified. For instance, in an external magnetic field, the dimer compound Yb$_2$Si$_2$O$_7$\cite{Hester2019} contains two intermediate phases between the dimerized state and PP, which is beyond the usual magnon BEC mechanism.  A previous theoretical work attributes the existence of two nontrivial intermediate phases to the anisotropy in the exchange interactions and the staggered Land\'{e} $g$ factor\cite{Flynn2021}.

On the one hand, in the layered honeycomb “Kitaev material” $\alpha$-RuCl$_3$ where SOC plays an important role\cite{Trebst2017,Motome2020,Senthil2020},
the dominant interactions are not AFM Heisenberg interactions but Kitaev interactions and other off-diagonal interactions.
At ambient pressure, the $\alpha$-RuCl$_3$ undergoes a first-order phase transition to a non-magnetic dimerized phase\cite{SunL2018,CuiYi2017,Brink2018,Huang2019}.
Another example is the previously mentioned dimerized honeycomb-lattice material Yb$_2$Si$_2$O$_7$ which may also have SOC-caused anisotropies since the Yb$^{3+}$ cation is very heavy.
This raises an interesting question: will the strong SOC affect the magnon condensation?
In the  present work, we will address this  question by studying the physical consequence of strong SOC to dimerized antiferromagnets. It will be illustrated that although the continuous symmetry is no longer a good approximation, {\it the magnetic-field-induced long-range AFM orders are quite general in dimerized antiferromagnets with SOC.} The 
magnetic order in the intermediate field region can be either N\'{e}el order, or stripe order, or zigzag order, depending on the details of interactions. Even when the ground state is N\'{e}el ordered state at zero field (owing to strong SOC), the in-plane stripe order or the zigzag order can still emerge in the intermediate field region.

We will begin with an antiferromagntic spin model on the distorted honeycomb lattice with finite SOC. It includes Heisenberg $J$ terms and off-diagonal $\Gamma'$ terms, namely,
\beq\label{JGM}
H \!\!& = &\!\! \sum_{\langle i,j \rangle \in \gamma}\!\! J_{\gamma}\pmb{S}_i\cdot \pmb{S}_j + \Gamma^{\prime}\left(S_{i}^{\alpha} S_{j}^{\gamma} \!+\! S_{i}^{\gamma} S_{j}^{\alpha} \!+\! S_{i}^{\beta} S_{j}^{\gamma} \!+\! S_{i}^{\gamma} S_{j}^{\beta}\right)   \nonumber\\
 & &-  g\mu_B\sum_i\pmb B\cdot\pmb{S}_i ,
\eeq
where $\langle i,j \rangle$ denotes nearest-neighbor sites, $\alpha, \beta, \gamma=x,y,z$ stand the spin indices, and $\gamma=x,y,z$ is also used to label the three types of bonds (see Fig.\ref{fig:hd}(a) for illustration). We adopt the setting with $g\mu_B=1$ and $J_z=J$, $J_x=J_y=(1-\delta) J$, $\delta \in [0,1)$. For simplicity, the magnetic field is oriented along c-direction.
With variational Monte Carlo (VMC) calculations using Gutzwiller projected states as trial wave functions, we found that field-induced 
magnetic orders, including N\'{e}el order, zigzag order, and stripe order, can appear in the intermediate magnetic fields (see Fig.\ref{fig:PD01} and Fig.\ref{fig:PD02} for phase diagrams). This suggests that SOC-caused frustration is crucial to understand the nature of the field-induced quantum critical points (QCP) and magnetic orders in the intermediate field region
(see Ref.~\onlinecite{YJW2016,Ye2017,Ye2020} for another example of model study).

The rest part of the paper is organized as follows.
In Sec.\ref{SecII}, we introduce the symmetry group of the extended Heisenberg model (the $J$-$\Gamma'$ model) on the honeycomb lattice. In Sec.\ref{SecIII}, we present the technical details of the VMC method, especially for the ground state and excited states. Readers who are not interested in the details of the method can skip Sec.\ref{SecIII} and go to Sec.\ref{SecIV} directly, where the VMC results are presented.
Sec.\ref{SecV} is devoted to the spin-wave theory and effective field theory aiming to understand the origin of the intermediate magnetically ordered phases and the nature of QCP. Finally, the paper is concluded in Sec.\ref{SecVI}.

\section{Symmetry group of the spin model}\label{SecII}

Generally, SOC in quantum magnets locks the spin rotation operations with the corresponding lattice rotations and hence reduce the continuous spin rotation symmetries to discrete ones. A class of spin-orbit coupled antiferromagnets on the honeycomb lattice has the discrete symmetry group $D_{3d}$\cite{Gegenwart2010,Kim2014,LiYuan2020,ZhangQingMing2021,Armitage2021} and contains anisotropic spin-spin interactions in addition to the isotropic Heisenberg exchanges ($J$ terms). The anisotropic exchange interactions include, the diagonal Kitaev exchange interactions $S_i^\gamma S_j^\gamma$ ($K$ terms)\cite{Kitaev2006,Khaliullin2009,Khaliullin2010} and the off-diagonal exchange interactions such as $S_i^\alpha S_j^\beta+S_i^\beta S_j^\alpha$ interactions ($\Gamma$ terms)\cite{Vojta2017,Moore2018,singleQ} or $S_i^\alpha S_j^\gamma+S_i^\gamma  S_j^\alpha + S_i^\beta  S_j^\gamma+S_i^\gamma S_j^\beta$ interactions ($\Gamma'$ terms)\cite{Rau2014, Valenti2016, Khaliullin2016, hidden2015}, here $\alpha,\beta,\gamma = x,y,z$ and we have only listed the interactions on the $\gamma$-bond.

On the other hand, some honeycomb-lattice materials are dimerized where the threefold rotational symmetry is absent. Supposing that the $z$-bonds are shorter than the other bonds, then it is expected that the strength of interactions is stronger on the $z$-bonds (see Fig.\ref{fig:hd}). In the present work, we will focus on the simple model (\ref{JGM}) on dimerized honeycomb lattice containing $J$ and $\Gamma'$ interactions only. Here we adopt the $\Gamma'$ term but do not consider other terms because the $\Gamma'$ interactions can give rise to a richer magnetic structure.  For simplicity, we assume that the strength of $\Gamma'$ interactions is equal on all the bonds, while the strength of the Heisenberg terms is anisotropic. Thus, we introduce a parameter $\delta$ to denote the relative difference of $J$ between the $z$-bonds and the $x,y$-bonds.
We apply the magnetic field along the $c$-direction with $\pmb B=B_c(1,1,1)/\sqrt 3$.

If $\delta=0, \pmb B=0$, the system has $D_{3d}\times Z_2^T$ symmetry\cite{Wang2019}, where $Z_2^T$ stands for the time reversal symmetry.
If $\delta\neq0, \pmb B=0$, the system has $\{E, \mathcal{P}, \mathcal{C}_2, \mathcal{M}\}\times Z_2^T$ magnetic point-group symmetry\cite{Wang2020}, with $\mathcal{P}$ the spatial inversion, $\mathcal{C}_2$ the twofold rotation along the $z$-bond,  and $\mathcal{M}=\mathcal{C}_2\mathcal{P}$ the mirror reflection symmetry.
For general values of $\delta$ and $\pmb B$ (e.g. $\delta\neq0, \pmb B\neq 0$), the symmetry of the model reduces to the following magnetic point group $2'/m'=\{E, \mathcal{P}, \mathcal{C}_2\mathcal{T}, \mathcal{M}\mathcal{T}\}$, with $\mathcal{T}$ the time reversal.

In the following, we will study the phase diagram and the nature of the phases for the spin model (\ref{JGM}) using VMC approach.

\section{The VMC approach}\label{SecIII}
\subsection{For the Ground States}\label{SecIIIA}

The VMC approach is based on spinon representation, where the spin operators are written in quadratic forms of fermionic spinons $S_i^m =\frac{1}{2} C_i^\dagger \sigma^m C_i$, where $C_i^\dagger = (c_{i\uparrow}^\dagger,c_{i\downarrow}^\dagger)$, $m \equiv x,y,z$, and $\sigma^m$ are Pauli matrices. The particle number constraint, $\hat{N_i} = c_{i\uparrow}^\dagger c_{i\uparrow} + c_{i\downarrow}^\dagger c_{i\downarrow} = 1$, should be imposed at every site such that the size of the Hilbert space of the fermions is the same as that of the original spin. It is convenient  to introduce the matrix operator $\psi_i=( C_i, \bar C_i)$ with $\bar C_i=(c_{i\dn}^\dag, -c_{i\up}^\dag)^T$ such that the spin operators can also be written as $S_i^m = {\rm Tr}(\psi_i ^\dag {\sigma^m \over4}\psi_i)$. Since the spin operator is invariant under a local SU(2) transformation $\psi_i\to \psi_i W_i$, the fermionic spinon representation has an SU(2) gauge symmetry\cite{Anderson88}.

The spin interactions are rewritten in terms of interacting fermionic operators and are further decoupled into a non-interacting mean-field Hamiltonian. Then we perform Gutzwiller projection to the mean-field ground state $|\Psi_{\rm mf} (\pmb R)\rangle$ to enforce the particle number constraint. The projected states $|\Psi (\pmb R)\rangle = P_G |\Psi_{\rm mf}(\pmb R) \rangle$ provide a series of trial wave functions depending on the choice of the mean-field Hamiltonian $H_{\rm mf}(\pmb R)$, where $P_G$ denotes a Gutzwiller projection and $\pmb R$ are treated as variational parameters.  The energy of the trial state $E (\pmb R) = \langle \Psi(\pmb R) |H| \Psi(\pmb R) \rangle / \langle \Psi(\pmb R)| \Psi(\pmb R) \rangle$ is computed using Monte Carlo sampling, and the optimal parameters $\pmb R$ are determined by minimizing the energy $E(\pmb R)$. In the following, we will focus on the construction of the trial mean-field Hamiltonian.

The Heisenberg interactions can be decoupled into a non-interacting fermionic Hamiltonian,
\beq
H_{\rm mf}^{J} &=& \sum_{\langle i,j \rangle \in \gamma} \left( t^{\gamma}C_{i}^{\dag}C_{j}+\Delta^{\gamma}C_{i}^{\dag}\bar{C_{j}} + {\rm H.c.} \right) \nonumber\\
&&
+\sum_{i} \operatorname{Tr}\left(\pmb{\lambda}_{i} \cdot \psi_{i} \pmb{\tau} \psi_{i}^{\dagger}\right),
\eeq
where $t^{\gamma}$ denotes singlet hopping and $\Delta^{\gamma}$ represents singlet pairing. $\lambda^{x,y,z}$ are three Lagrangian multipliers to ensure the SU(2) gauge invariance (where the $\lambda^z$ component corresponds to the particle number constraint).

To decouple $\Gamma'$ interactions in the model (\ref{JGM}),  we adopt the “Kitaev-type” mean-field Hamiltonian appeared in literature\cite{Wang2019,Gamma'2020,Wang2020,LiHan2021}, namely,
\beq\label{Gmp}
H_{\rm mf}^{\Gamma'}\!
& = &  \sum_{\langle i,j \rangle \in\gamma} \!\! i \rho_c {\rm Tr} \left(\psi_i^\dagger \psi_j + \tau^x \psi_i^\dagger
\sigma^x \psi_j + \tau^y \psi_i^\dagger \sigma^y \psi_j \right. \nonumber \\ & &
\;\;\; \left. + \tau^z \psi_i^\dagger \sigma^z \psi_j \right) + i\rho_b {\rm Tr} \left( \tau^\alpha \psi_i^\dag
\sigma^\gamma \psi_j + \tau^\gamma \psi_i^\dag
\sigma^\alpha \psi_j \right. \nonumber \\ & &
\;\;\; \left.  + \tau^\beta \psi_i^\dag \sigma^\gamma \psi_j + \tau^\gamma \psi_i^\dag \sigma^\beta \psi_j \right) +  {\rm H.c.}
\eeq
where $\rho_b$ and $\rho_c$ are both real numbers.
Note that $\tau^{x,y,z}$ are generators of the SU(2) gauge group\cite{Wang2019}. Meanwhile, the mean-field Hamiltonian $H_{\rm mf}^{\Gamma'}$ preserves all symmetries of the original system through projective symmetry group\cite{Wen2002}.

The Zeeman coupling is simply written as
\beq\label{Bfield}
H^{\rm B} = - \sum_i {\rm Tr}  \left(\pmb B\cdot  \psi_i^\dag {\pmb \sigma\over2}\psi_i\right).
\eeq

Finally, to describe the field-induced magnetic order in the $J$-$\Gamma'$ model, we introduce a background field $\pmb M_i$ to induce the symmetry-breaking magnetic order. The ordering pattern $\pmb M_i$ is assumed to contain a single momentum $\pmb Q$ with\cite{singleQ}
$$\pmb{M}_i = M \Big( \sin \phi \big[\hat {\pmb e}_x \cos (\pmb{Q} \cdot
\pmb{r}_i) + \hat{\pmb e}_y \sin(\pmb{Q} \cdot \pmb{r}_i)\big] + \cos \phi
\, \hat{\pmb e}_z \Big ),$$
where  $\pmb Q$ is the ordering momentum, $\hat {\pmb e}_{x,y,z}$ are the local spin axes, and $\phi$ is the canting angle. The ordering momentum $\pmb Q$ is adopted either from the classical ground state or the classical metastable states (the $\pmb Q$ from the classical ground state is not always the one with the lowest energy after quantum corrections are considered). For a given $\pmb Q$, the local axes $\hat {\pmb e}_{x,y,z}$ are fixed as they are in the classical state, while $M$ and $\phi$ are treated as variational parameters.

Hence, the complete trial mean-field Hamiltonian (with nearest-neighbor coupling terms only) for the $J$-$\Gamma'$ model in an external magnetic field reads
\begin{equation}\label{Order}
H_{\rm mf}^{\rm total} = H_{\rm mf}^{J} +H_{\rm mf}^{\Gamma'} +H^{B} - {\textstyle \frac{1}{2}} \sum_i
(\pmb {M}_i \cdot C_i^\dagger \pmb \sigma C_i + {\rm H.c.}).
\end{equation}

The complete set of variational parameters include ($t^\gamma$, $\Delta^\gamma$, $\rho_b$, $\rho_c$, $\pmb\lambda$, $M$, $\phi$) whose optimal values are determined by minimizing the energy of the Gutzwiller projected wave function.

In the variational process, we first choose a classical metastable configuration with $\pmb Q$, then optimize the energy of the projected state. By comparing the optimal energies of different trial $\pmb Q$s, we obtain the approximate ground state.

\begin{figure*}[t]
\includegraphics[width=5.9cm]{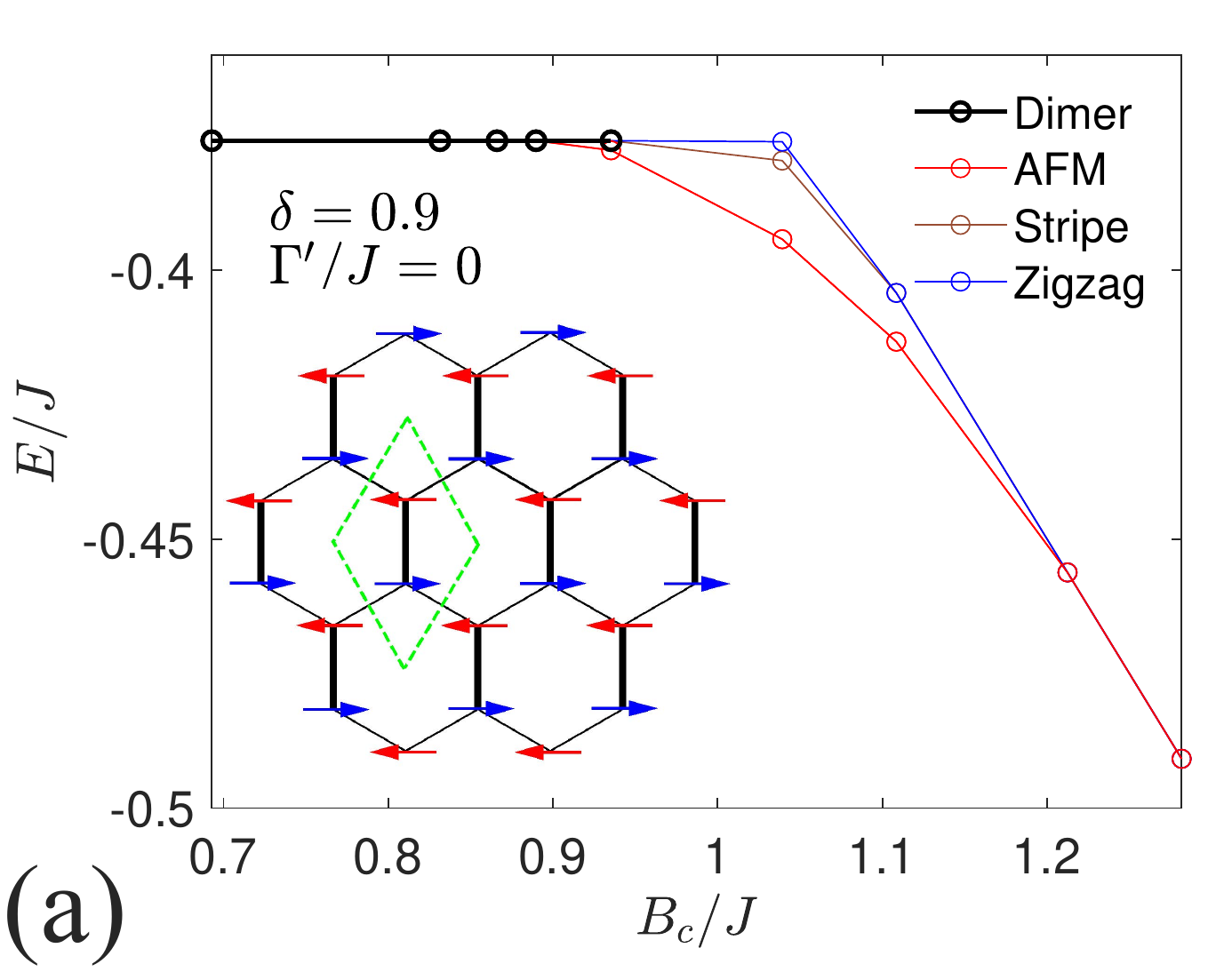}\!
\includegraphics[width=5.9cm]{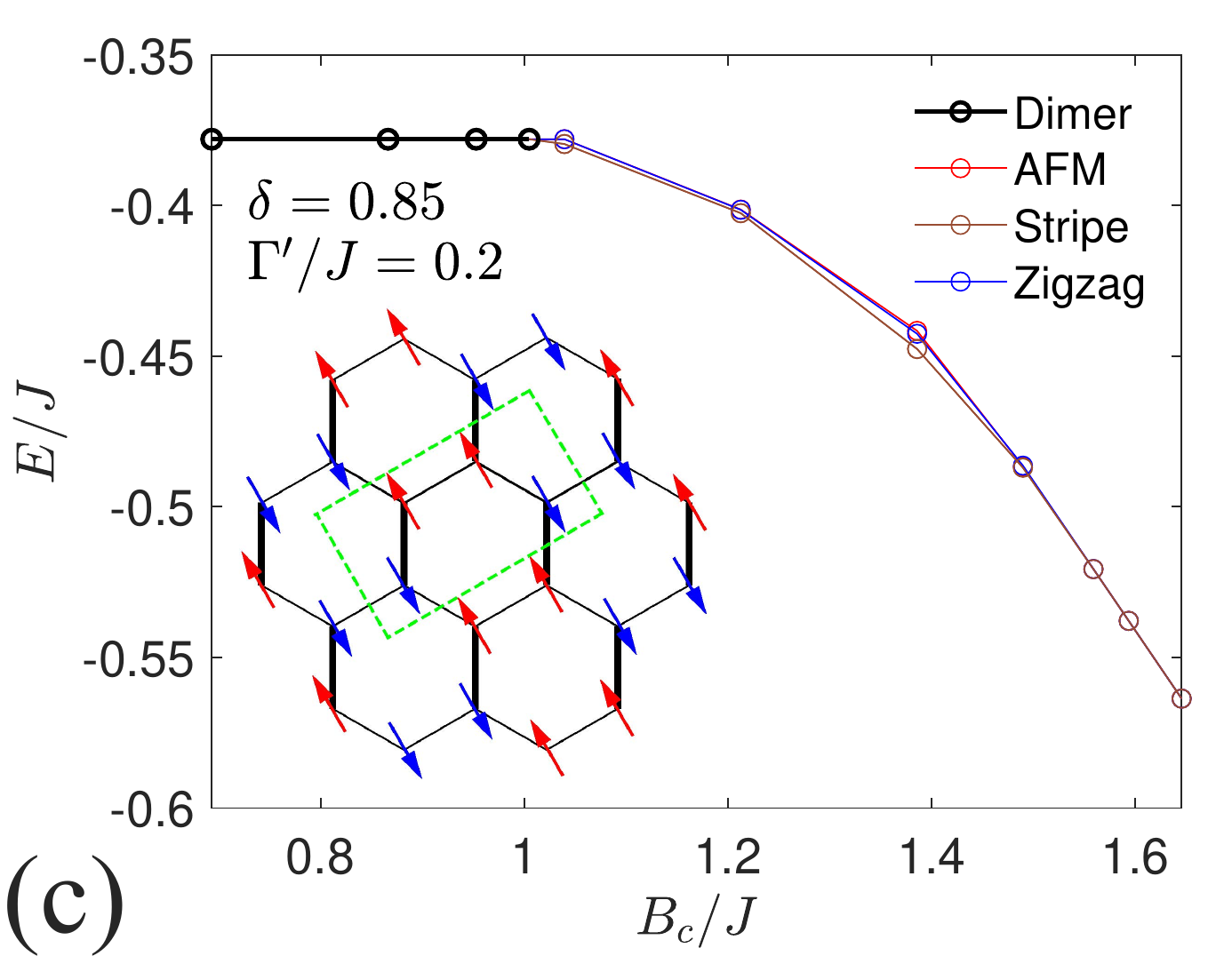}\!
\includegraphics[width=5.9cm]{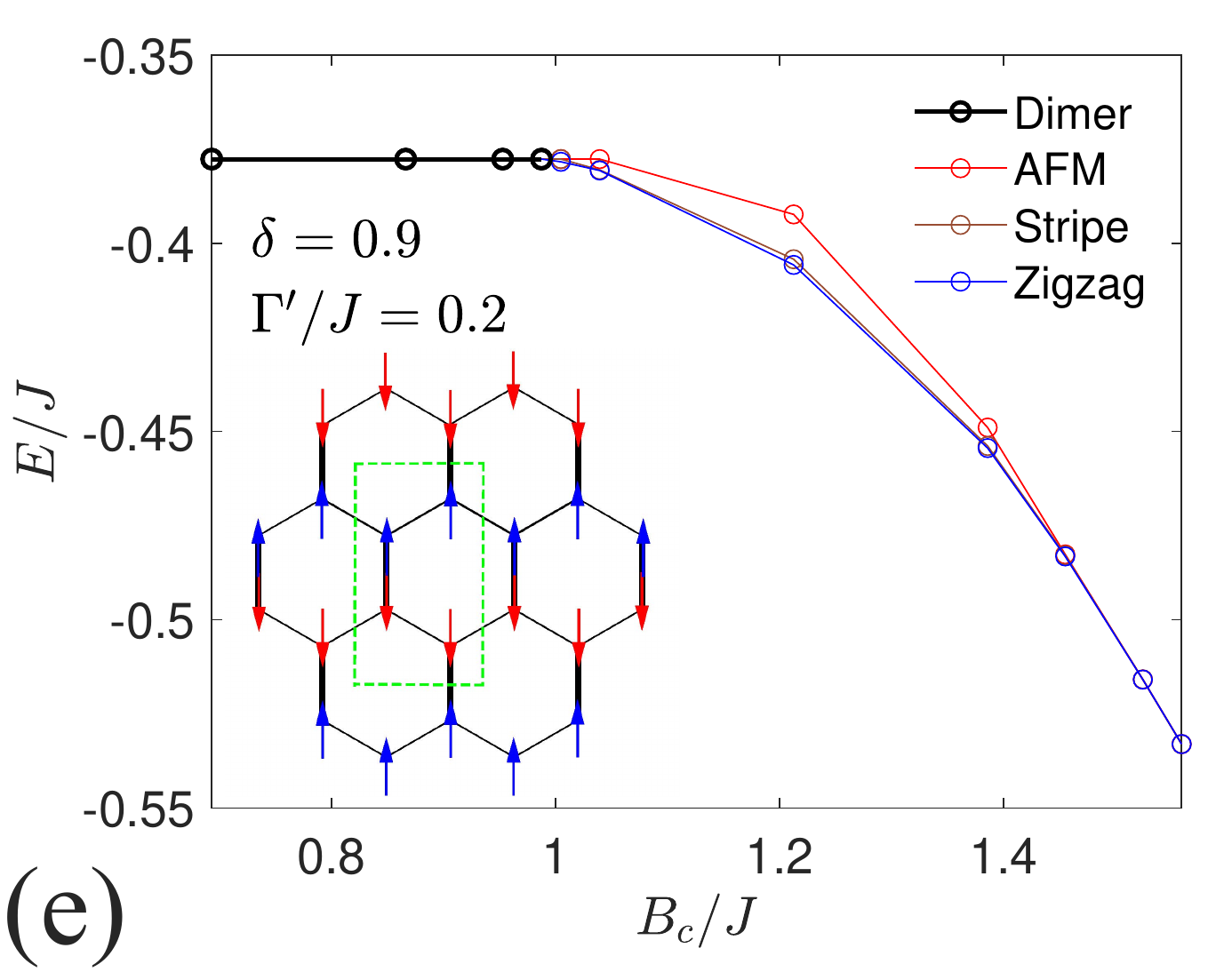} \\
\includegraphics[width=5.8cm]{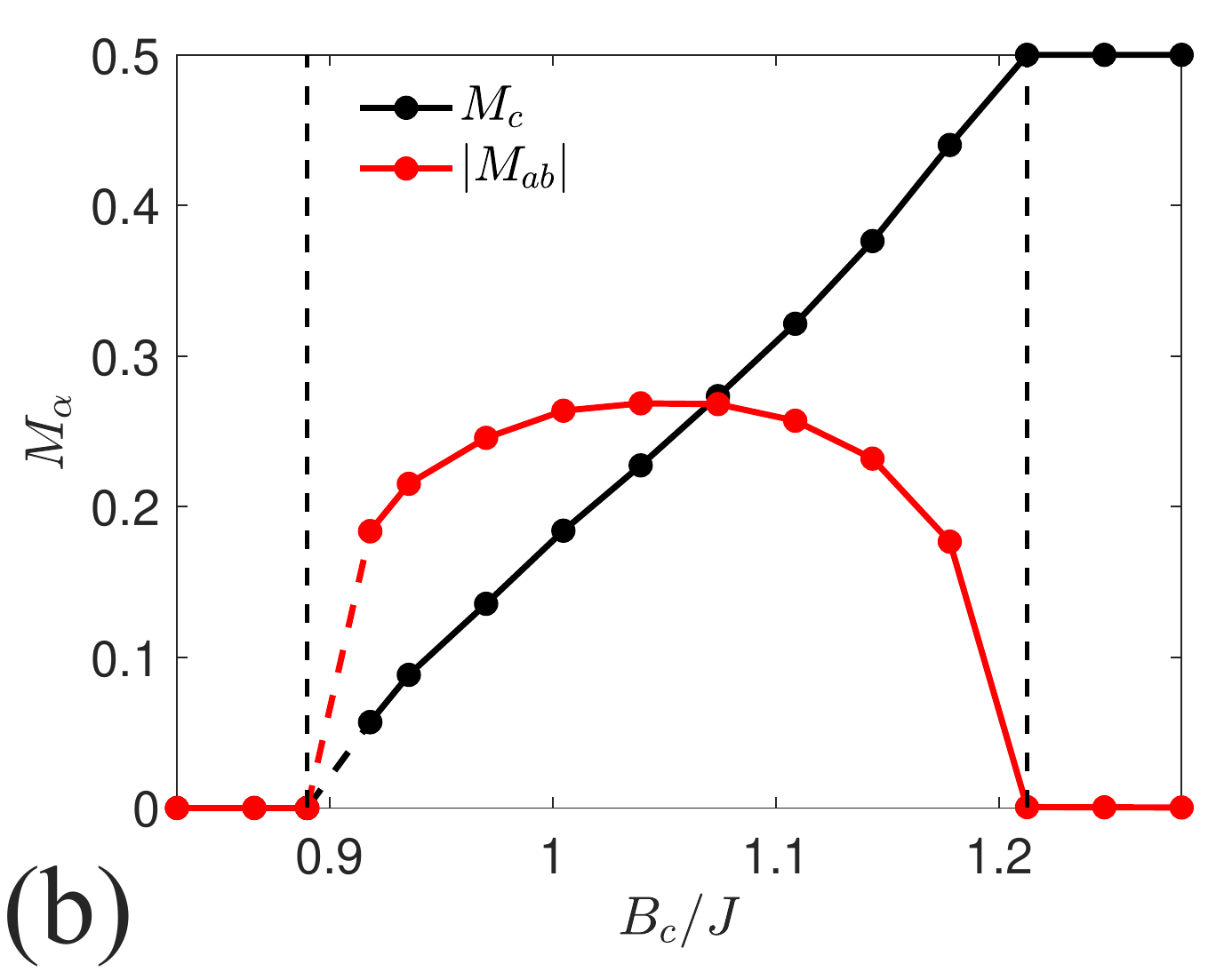}\!
\includegraphics[width=5.8cm]{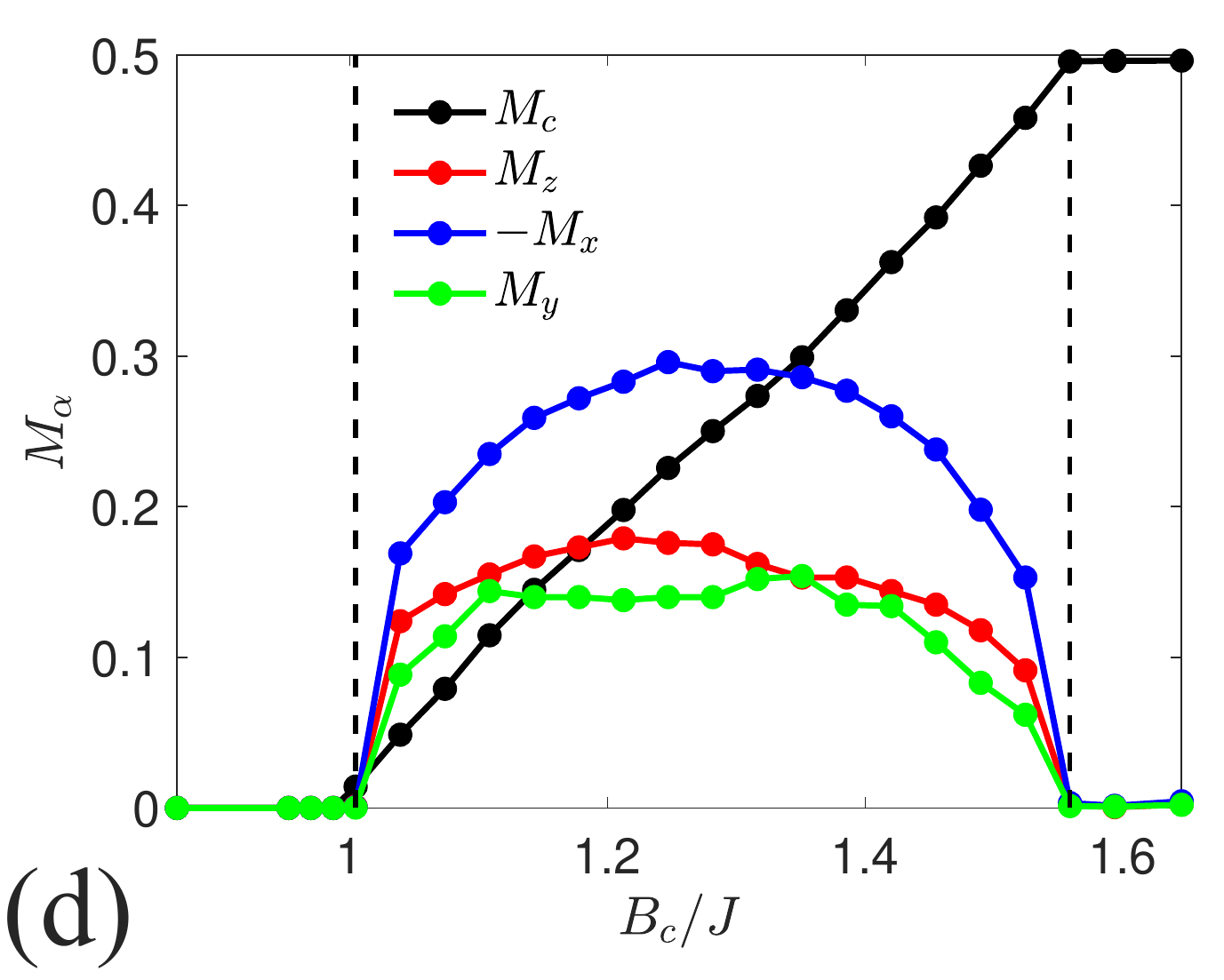}\!
\includegraphics[width=5.8cm]{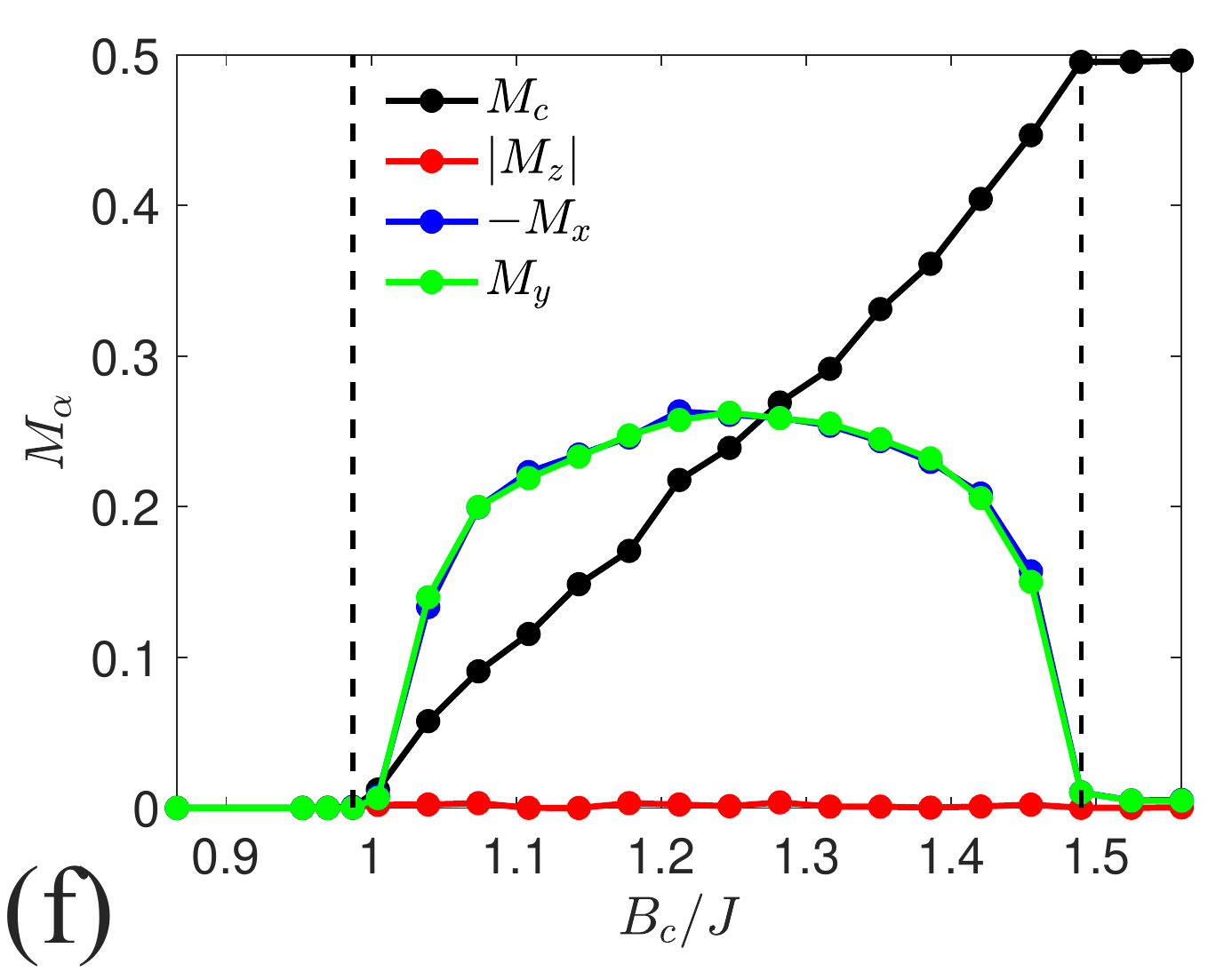}\!
\caption{
Energy curves (the upper figures) and magnetization of ground states (the lower figures) with increasing magnetic field for (a)\&(b) $\delta$=0.9, $\Gamma^{\prime}/J$=0, (c)\&(d) $\delta$=0.85, $\Gamma^{\prime}/J$=0.2, (e)\&(f) $\delta$=0.9, $\Gamma^{\prime}/J$=0.2, which are points in region-I, region-II, and region-III with star [see Fig.\ref{fig:PD01}(b)], respectively. The insets of (a), (c), (e) show the spin configurations of AFM, stripe, and zigzag orders, respectively. The vertical dashed lines indicate the values of the critical fields.
}\label{fig:EM}
\end{figure*}

In addition to determining the phase diagram (see Fig.\ref{fig:PD01}, Fig.\ref{fig:PD02}, and Fig.\ref{fig:EM}), the VMC method can further provide information on the low-energy excitations.

\subsection{For the Spinon-Excited States }\label{SecIIIB}

Now we try to construct the low-energy excitations via Gutzwiller projected spinons excited states
\beq\label{spinonpq}
|\pmb k;({\pmb k - \pmb q})_m,({\pmb q})_n\rangle = P_G f_{\pmb k-\pmb q,m}^\dag f_{\pmb q,n}^{\dag} |\Psi_{\rm mf} \rangle,
\eeq
where $f_{\pmb k-\pmb q,m}$ and $f_{\pmb q,n}$ are eigen particles of the mean-field Hamiltonian, with $m,n$ the band indices and $\pmb k,\pmb q$ the lattice momentum according to the translation operators.

The above two-spinon excitations form a continuum and do not correctly describe the low-energy excitations of the system. The failure of the state (\ref{spinonpq}) in describing  the low-energy excitations is owing to the strong gauge interactions between the spinons which are not correctly addressed. To partially solve this problem, we diagonalize the original Hamiltonian in the subspace spanned by the two-spinon excitation continuum\cite{TaoLi2010,FanYang2011,Monnow2015,Zhao2021}. specifically, we calculate the matrix elements of the matrix $\mathscr H(\pmb k)$ with
\Beq
\mathscr H_{pmn,qm'n'}(\pmb k) = \langle  \pmb k; (\pmb k-\pmb p)_m, (\pmb p)_n |H| \pmb k; (\pmb k-\pmb q)_{m'}, (\pmb q)_{n'} \rangle,
\Eeq
where $\pmb k$ is the total momentum of the state with a pair of spinons excitation. The dimensionality ($2N\times C$) of this matrix $\mathscr H(\pmb k)$ is equal to twice the total number of sites $N$ times the size of the unit cell $C$ in the corresponding mean-field Hamiltonian.

Since the states in the two-spinon continuum are not orthogonal, we need to calculate the metric matrix $g$ formed by the overlap of the states,
\[
g_{pmn,qm'n'}(\pmb k) =\langle  \pmb k; (\pmb k-\pmb p)_m, (\pmb p)_n | \pmb k; (\pmb k-\pmb q)_{m'}, (\pmb q)_{n'} \rangle.
\]
Hence, the eigenproblem of $\mathscr H(\pmb k)$ should be calculated by
\[
 g^{-1}(\pmb k)\mathscr H(\pmb k) U = U\cdot {\rm diag}(\epsilon_1, ...., \epsilon_{2CN}),
\]
where the eigenvalues $\epsilon_1, ...., \epsilon_{2CN}$ are the `renormalized' energy of the excitations.

The renormalized eigenfunction $|\pmb k\rangle_{\rm rn}$ is given by
\Beq
|\pmb k\rangle_{\rm rn} = \sum_{\pmb q\in BZ} \mathscr F(qmn)|\pmb k; (\pmb k-\pmb q)_m,(\pmb q)_n\rangle,
\Eeq
where $\mathscr F(qmn)$ is the eigenvector of the matrix $\mathscr H$.

It will be shown that in the magnetically ordered phase the two-spinon excitations are strongly renormalized and the bound state formed by the spinons pair behaves like a magnon (see Fig.\ref{fig:Excited_DP} and Fig.\ref{fig:DSF_J1_0d9_Bz0d58}).

\section{The  VMC results}\label{SecIV}

\subsection{The Phase Diagram}

Our VMC calculations are performed on a  tori of 6$\times$6 unit cells, {\it i.e.}~of 72 lattice sites. The main results are shown in Fig.\ref{fig:PD01}. The ($\Gamma'$,$\delta$)-parameter space is divided into several different regions. Firstly, when $\pmb B=0$, the phase diagram contains two phases that are separated by the solid line. The lower one is a QDM, while the upper one is a N\'{e}el ordered state locating perpendicular to the honeycomb plane (namely, AFM$_c$). The  phase transition between both gapped phases is of second order with a dynamical critical exponent $z=1$ (see Appendix \ref{app:1stdim} for details) which is characterized by the  spontaneous breaking of spatial inversion symmetry.

When $\pmb B\neq0$, according to the pattern of intermediate magnetic orders at intermediate magnetic fields, each phase is divided into three different regions [see Fig.\ref{fig:PD01}(b) for details].

In region-I, the $\Gamma'$ interaction is relatively small, the SOC is not strong enough to change the physics of magnon BEC. Namely, there is an intermediate canted AFM with long-range magnetic order with N\'{e}el pattern in the plane perpendicular to the field (namely, AFM$_{ab}$) between the low-field QDM and the high-field PP [see Fig.\ref{fig:EM}(a) for illustration], just as the case $\Gamma'=0$.
The magnon excitations are gapless in the intermediate AFM$_{ab}$ phase although $\Gamma' \neq0$, on account of a continuous U(1) degenerate manifold of ground states in the corresponding classical model\cite{Kim2020}. Quantum fluctuations lifts the U(1) degeneracy and leads to the pseudo-Goldstone mode, which is known as order-by-disorder effect\cite{Henley1989}.
It can also be seen that either the canted stripe order or canted zigzag order with long-range magnetic order in the plane perpendicular to the field (namely, stripe or zigzag) is competing in energy with the N\'{e}el state.
With the increasing of field strength, the phase transitions into the AFM$_{ab}$ phase and out of AFM$_{ab}$ phase are both of second order, which is the same as the case with $\Gamma'=0$.  Here we clarify the fact that the magnetization of the ground state has a small jump at the low-field transition point, which seems to indicate a weakly first-order transition.
We speculate that this is an artifact owing to the VMC method, where the magnetic order is induced by a static background field. We could resolve the problem through ‘renormalization’ in VMC framework that allows the ground state to include more spin-wave fluctuations, which will smoothen the magnetization curve.
Meanwhile, the nature of the phase transitions will be analyzed in the next section using the renormalized spinons excitations.

\begin{figure*}[t]
\includegraphics[width=5.6cm,height=4.0cm]{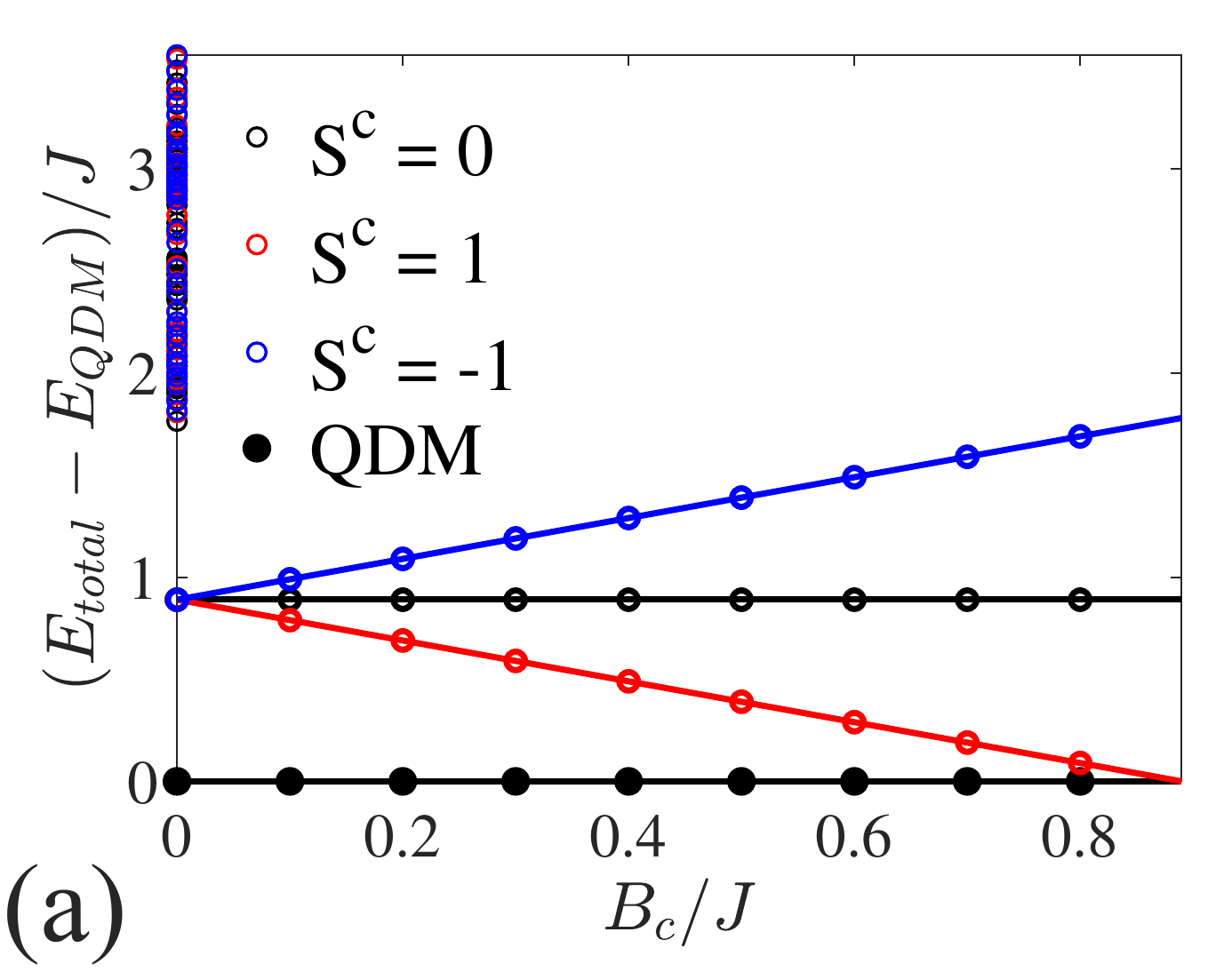}\!
\includegraphics[width=5.6cm,height=3.9cm]{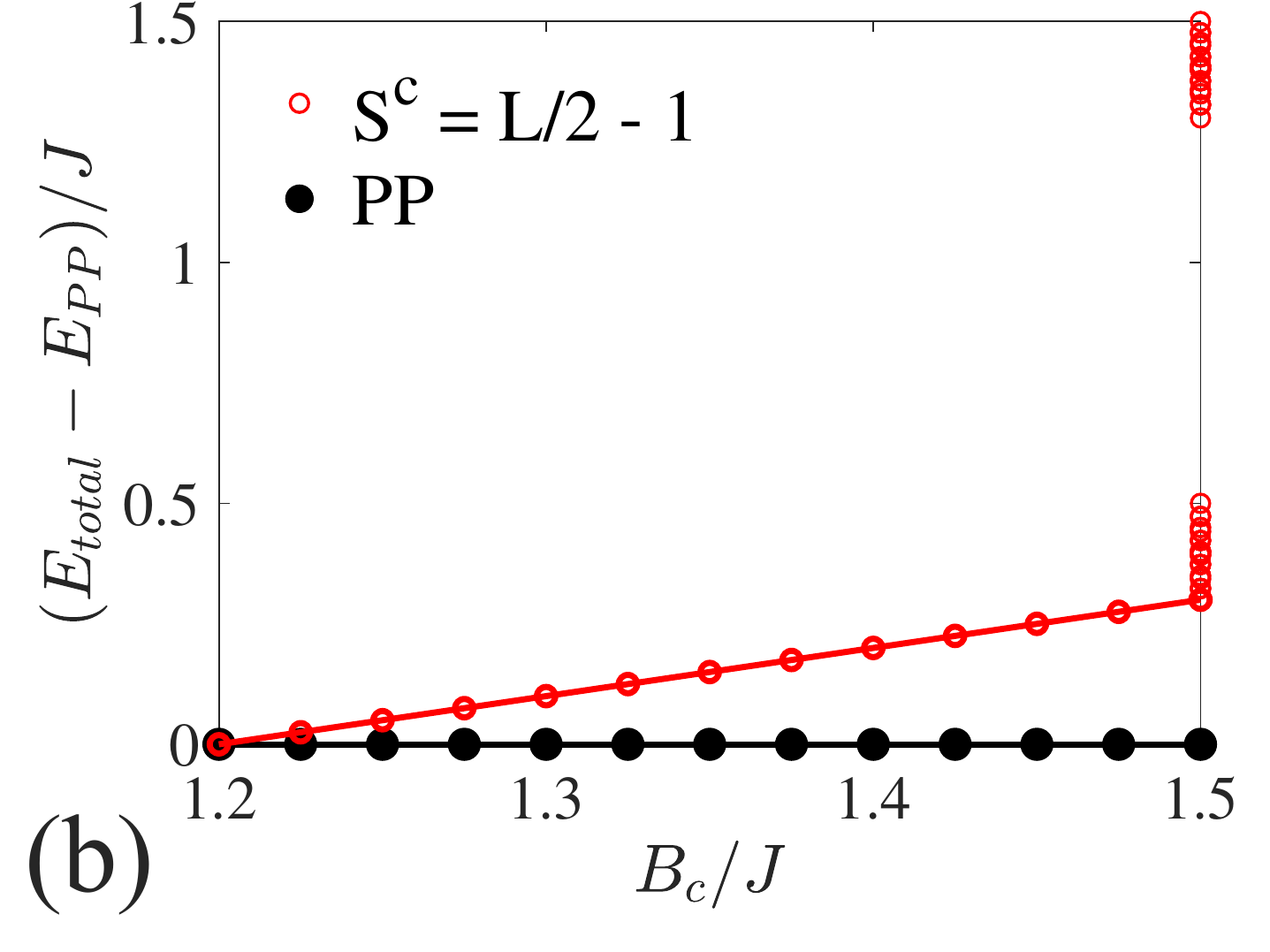}\!
\includegraphics[width=5.6cm,height=3.9cm]{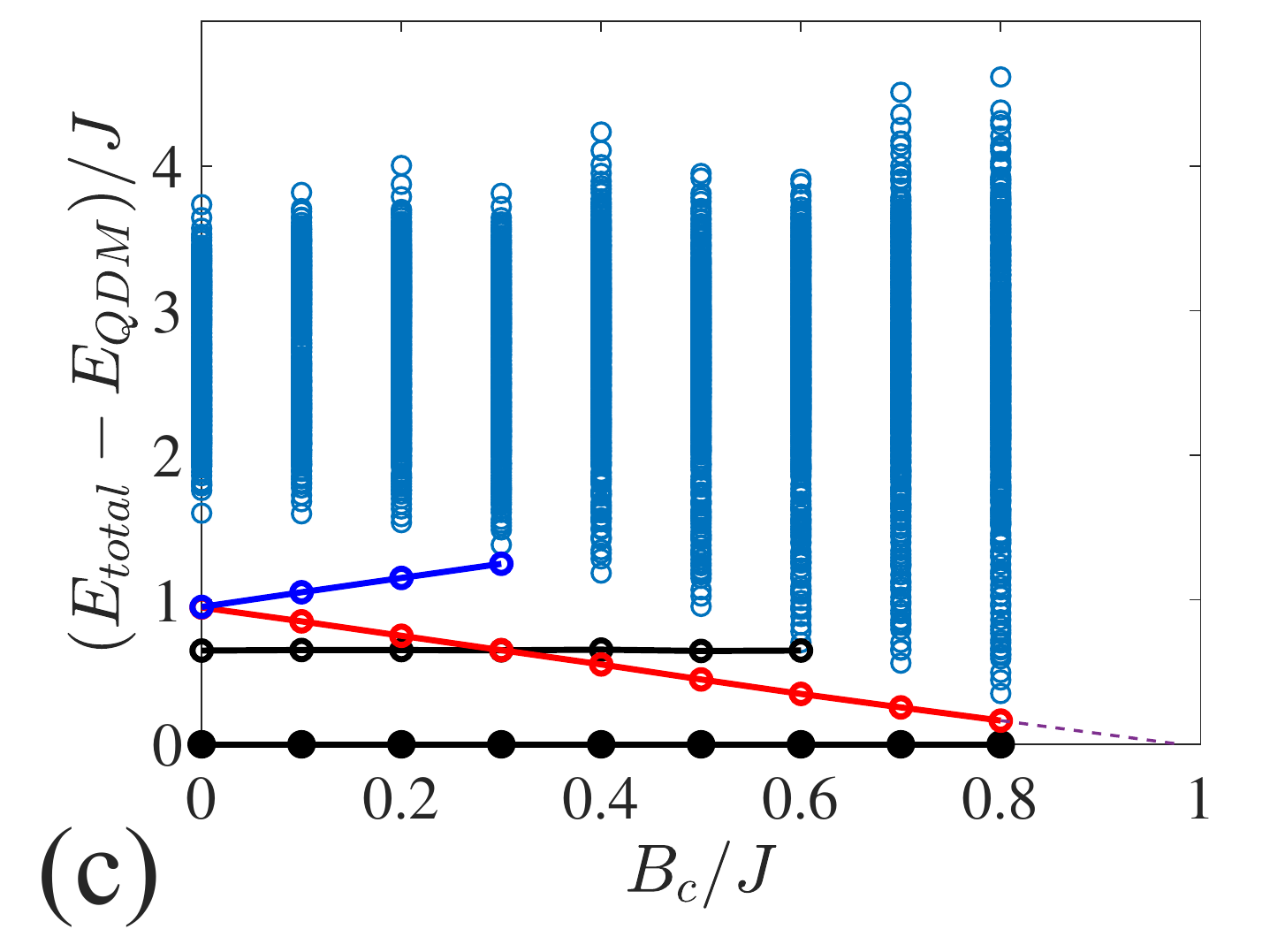}
\caption{ Energy spectrum at $\pmb \Gamma$ point ($\pmb k=0$) as a function of magnetic field strength, where the interaction parameters are fixed as $\delta=0.9$, $\Gamma'/J=0$ for (a)\&(b) and $\delta=0.8$, $\Gamma'/J=0.1$ for (c). (a) VMC is used in the QDM phase. At $B_c=0$, the system has a spin-singlet ground state, gapped triplet excitations (with $S^c=0,\pm1$) owing to the continuous SU(2) spin rotational symmetry and higher-energy continuum spectra. All these states are unaffected by the  field below $|B_{c}/J|<0.89$ but the energy changes linearly with $B_c$ until the gap closes. (b) Exact diagonalization in $S^c=L/2-1$ space is used in the polarized limit. As $B_c/J \geq 1.2$, the ground state is fully polarized with $S^c=L/2$, while the excited states with $S^c=L/2-1$ form a continuum with a gap above the ground state. These states are unaffected by the field but the energy changes linearly with $B_c$ until the gap closes at $B_c/J=1.2$.
(c) VMC is used in the QDM phase with finite SOC. The lowest energy excitation states at $B_c=0$ are not degenerate because of explicit breaking of SU(2) symmetry. However, the field dependence of the energy lines marked with black, blue, red  are very similar to the case in (a) where the SO(2) rotation symmetry is preserved.
}\label{fig:Excited_DP}
\end{figure*}

With the increasing of $\Gamma'$ and $\delta$, the region-II shows up. The difference is that in the intermediate phase, the stripe ordered state is slightly lower in energy than the N\'{e}el state and the zigzag state [see Fig.\ref{fig:EM}(c) for illustration].  In the stripe ordered phase, the spin components in the honeycomb plane are pointing along [$\overline{2}$11] direction [see Fig.\ref{fig:EM}(d)], although the canting direction is slightly away from c-direction. Among the symmetry elements $2'/m'=\{E, \mathcal{P}, \mathcal{C}_2\mathcal{T}, \mathcal{M}\mathcal{T}\}$ of the Hamiltonian, the $\mathcal{C}_2\mathcal{T}$ and $\mathcal{M}\mathcal{T}$ symmetries are broken by the stripe order.
The magnon excitations of the stripe phase are gapped, which is different from the case in AFM$_{ab}$ state.
Our numerical simulations strongly suggest that the transition between the intermediate stripe state and the dimerized (or polarized) phase is continuous, which is in agreement with its spontaneous symmetry breaking.

With further increasing of $\delta$, the system enters the region-III, where the intermediate phase exhibits zigzag type magnetic order [see Fig.\ref{fig:EM}(e) for illustration]. The in-plane zigzag order is orienting along the [1$\overline{1}$0] direction [see Fig.\ref{fig:EM}(f)], which breaks the $\mathcal{C}_2\mathcal{T}$ symmetry but preserves the $\mathcal{M}\mathcal{T}$ symmetry.
The magnon excitations of the zigzag phase are also gapped.
The phase transition between the intermediate zigzag state and the dimerized (or polarized) phase is also continuous owning to spontaneous symmetry breaking, which is the same as the case in region-II.

\begin{figure*}[t]
\includegraphics[width=8.3cm]{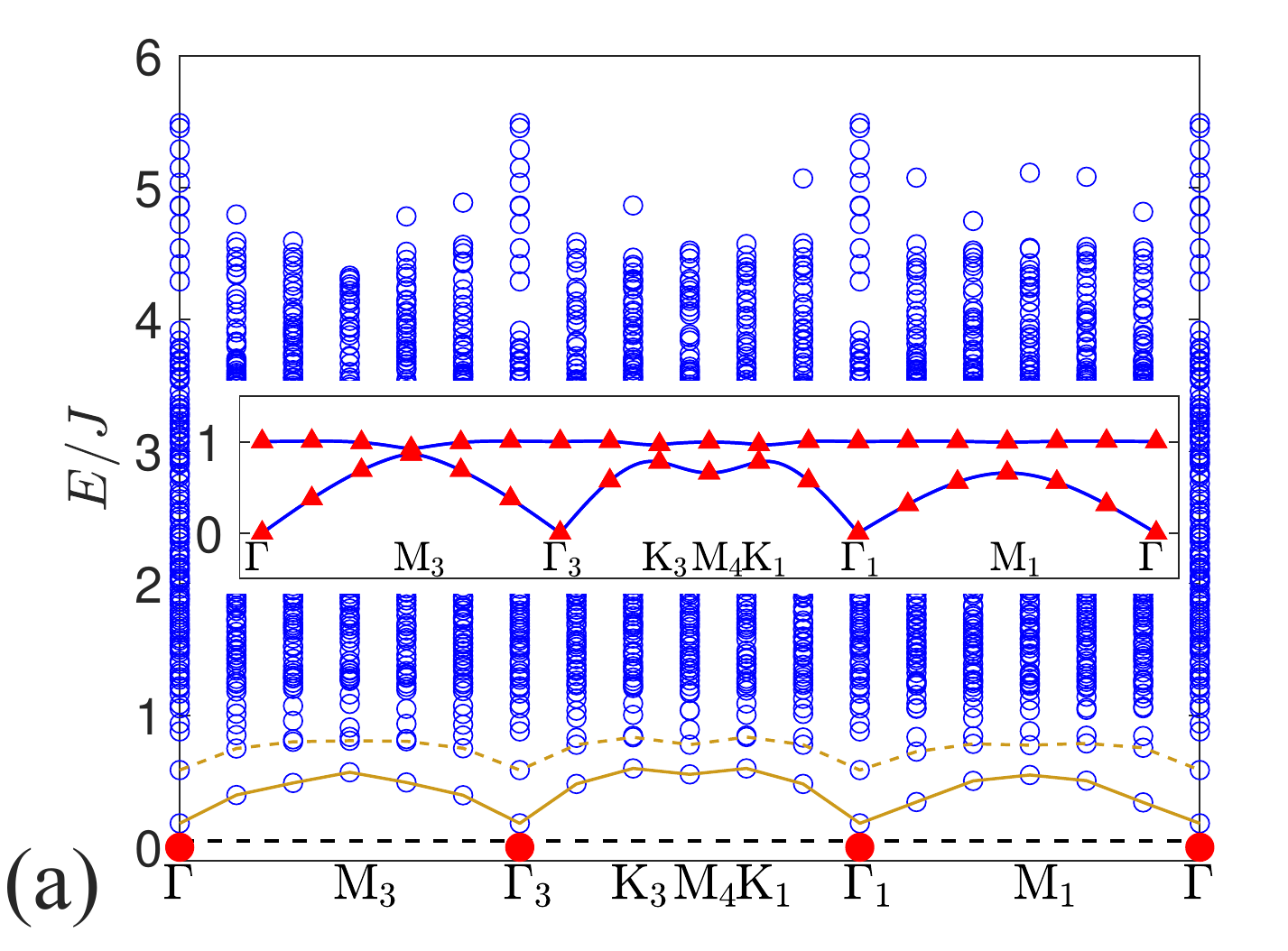}\!
\includegraphics[width=8.3cm]{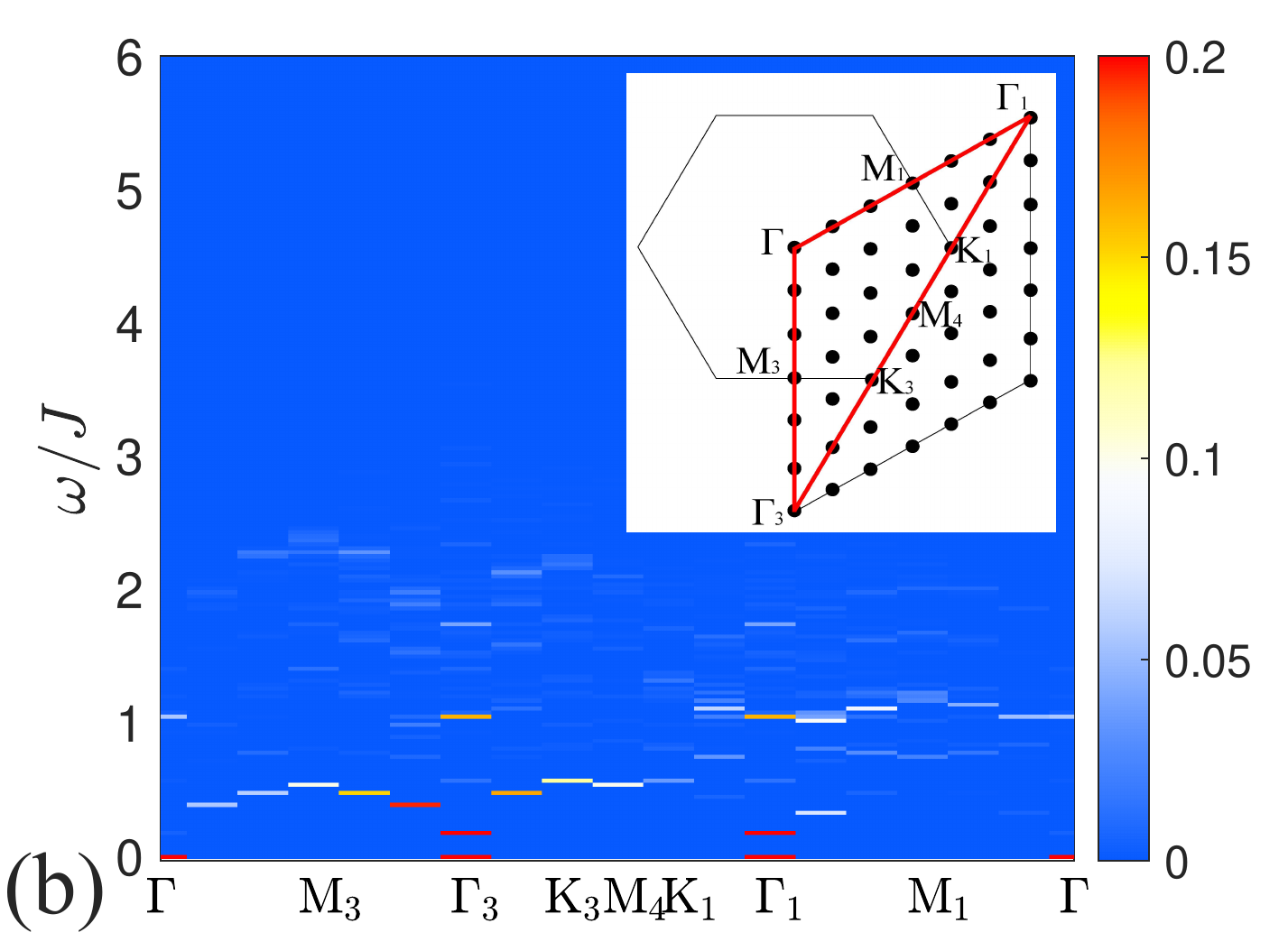}\!
\caption{(a) VMC excitation spectrum and (b) dynamic structure factor for the $\rm AFM_{ab}$ phase (in a system with 72 sites) with parameters $\delta=0.6$, $\Gamma'/J=0$, $B_c/J=1$. The black dashed line in (a) is the energy of the ground state before the `renormalization', while the red dot stands for the `renormalized' one. The inset of (a) shows the one-magnon bands from LSWT, and the inset of (b) illustrates the BZ and the path of the excitation spectrum.
}\label{fig:DSF_J1_0d9_Bz0d58}
\end{figure*}

In the left up region, owing to the relatively large $\Gamma'$, the zero-field ground state 
is no longer QDM with singlet dimers but is replaced by an AFM$_c$ state.
Among the symmetry elements $2'/m'=\{E, \mathcal{P}, \mathcal{C}_2\mathcal{T}, \mathcal{M}\mathcal{T}\}$ of the Hamiltonian, the $\mathcal{M}\mathcal{T}$ and $\mathcal{P}$ symmetries are broken by the AFM$_c$ order at low fields.
According to the response to the magnetic field, this region is further divided into three parts.
In the first part, with the increase of magnetic field, there is an intermediate AFM ordered state (with N\'{e}el pattern in the honeycomb plane, called AFM$_{ab}$) under a proper magnetic field as in the region-I [see Fig.\ref{fig:PD01}(b) and Fig.\ref{fig:PD02}(I)\&(IV)].
In the second part, there is an intermediate stripe phase under a proper magnetic field as in the region-II [see Fig.\ref{fig:PD01}(b) and Fig.\ref{fig:PD02}(II)\&(V)].
In the last part, there is also an intermediate phase that exhibits zigzag order as in the region-III [see Fig.\ref{fig:PD01}(b) and Fig.\ref{fig:PD02}(III)\&(VI)].
The magnon excitations are also gapped in the intermediate stripe and zigzag phases.
In all the above cases, the phase transition from the intermediate phase to the dimerized or polarized phase is continuous, while the transition from the intermediate phase to the low-field AFM$_c$ phase is of first order (for details see Appendix \ref{app:1st}).

From the phase diagram, we can see that the dimerized ground state 
is not a necessary starting point 
for the existence of the field-induced intermediate magnetically ordered phase. Therefore, the physics of this system would be further enriched by proper SOC.

\subsection{Excitations and Dynamical Properties} 

 In this section, we will focus on the low-energy excitations from the VMC approach. We first illustrate the closure of magnon excitation gap, which causes the condensation of magnons,  and then show the dynamic structure factor of the intermediate phase.

\subsubsection{Low-field and high-field limit}

As a typical example, we mainly focus on the interaction parameter region of the dimerized phase.
In the dimerized phase, the spinons are confined in the singlet ($S=0$), three energy-degenerate triplets ($S=1$), and high-energy continues.

The renormalized excitation spectrum for $\pmb \Gamma$ point are shown in Fig.\ref{fig:Excited_DP}(a). Here we can clearly see the singlet and triplet states at low energies, and higher energy continues. Because SOC disappears here, $S^c$ along the magnetic field (i.e. $c$-direction) is a good quantum number, then we could divide them into three classes according to $S^c= -1,0,1$. When a magnetic field is added, Zeeman terms are added. In orthogonal normalization excitation space, Zeeman terms only appear in diagonals, and produce a splitting of energy without a change of state. Energy splitting lines between singlet and triplet states are plotted in Fig.\ref{fig:Excited_DP}(a), they close the gap at $B_c/J \approx0.89$, which means a continuous transition, i.e. magnon BEC.

At the high-field limit, the polarized state is the ground state, and $S^c=L/2$ is also a conserved quantity. The space of low-energy excited states is $S^c=L/2-1$ with dimension $L$, hence we could obtain exact Hamiltonian $\mathcal{H}$ in this space, $\mathcal{H}_{i,j} = \langle i|H|j  \rangle$, where $|i(j)\rangle$ means the configuration flip i(j)-th from the polarized state. Then diagonalize $\mathcal{H}$ to get the excitation spectrum with various fields, as shown in Fig.\ref{fig:Excited_DP}(b). When the magnetic field drops to $B_c/J =1.2$, the gap closing also indicates a continuous phase transition from the polarized phase to the AFM$_ab$ phase.
These results are consistent with previous studies\cite{Giamarchi2008, Zapf2014}, which once again demonstrate the reliability of our VMC method.
If SOC appears (namely, nonzero $\Gamma'$ interactions), similar results will indicate continuous phase transitions as shown in Fig.\ref{fig:Excited_DP}(c).

\subsubsection{Dynamical structure factor in the intermediate phase}

Now we study the low-energy physics of the intermediate phase and will focus on the $\rm AFM_{ab}$ phase as an example. We will calculate the excitation spectrum and the dynamic structure factor (DSF)\cite{BeccaDSF} with interaction parameters $\delta=0.6, \Gamma'/J=0, B_c/J=1$.

The excitation spectrum and DSF for the $\rm AFM_{ab}$ phase are obtained from the renormalized spinons excitations as discussed in Sec.\ref{SecIIIB}. For simplicity, we consider the path in momentum points as plotted in the inset of Fig.\ref{fig:DSF_J1_0d9_Bz0d58}(b), which includes 2 different $\pmb K$ points and 3 different $\pmb M$ points.
The magnons as bound states of spinons are clearly seen in Fig.\ref{fig:DSF_J1_0d9_Bz0d58}(a), which are well separated from the continuum.
Especially, the lowest energy band (the orange solid line) agrees well with the lowest magnon band calculated from the linear spin-wave theory (LSWT) [see the insert of Fig.\ref{fig:DSF_J1_0d9_Bz0d58}(a)]. The small gap at the $\pmb \Gamma$ point in the magnon is owing to the system size effect. However, the second excitation band [the orange dashed line in Fig.\ref{fig:DSF_J1_0d9_Bz0d58}(a)] is dispersive and deviates with the dispersion of the upper magnon band (which is quite flat) in LSWT.

Fig.\ref{fig:DSF_J1_0d9_Bz0d58}(b) shows the  $zz$-component of the DSF which can be measured in neutron scattering experiments,
\Beq
S^{zz}(\pmb q, \omega) = \sum_{n}\left|\left\langle\Psi_{n}^{\pmb q}|S_{\pmb q}^{z}|\Psi_{0}\right\rangle\right|^{2} \delta\left(\omega-E_{n}^{\pmb q}+E_{0}\right)
\Eeq
where $\left|\Psi_{0}\right\rangle$ is the variational ground state with energy $E_{0}$ and $\left|\Psi_{n}^{\pmb q}\right\rangle$ is the n-th `renormalized' two-spinon excited state with momentum $\pmb q$ and energy $E_{n}^{\pmb q}$. We note that $S_{\pmb q}^{z}=\frac{1}{\sqrt{L}} \sum_{\pmb r} \exp [i \pmb q \cdot \pmb r] S_{\pmb r}^{z}$ is the Fourier-transformed spin operator for the component $z$, with $L$ being the number of sites in the system.

\begin{figure*}[t]
\includegraphics[width=5.7cm]{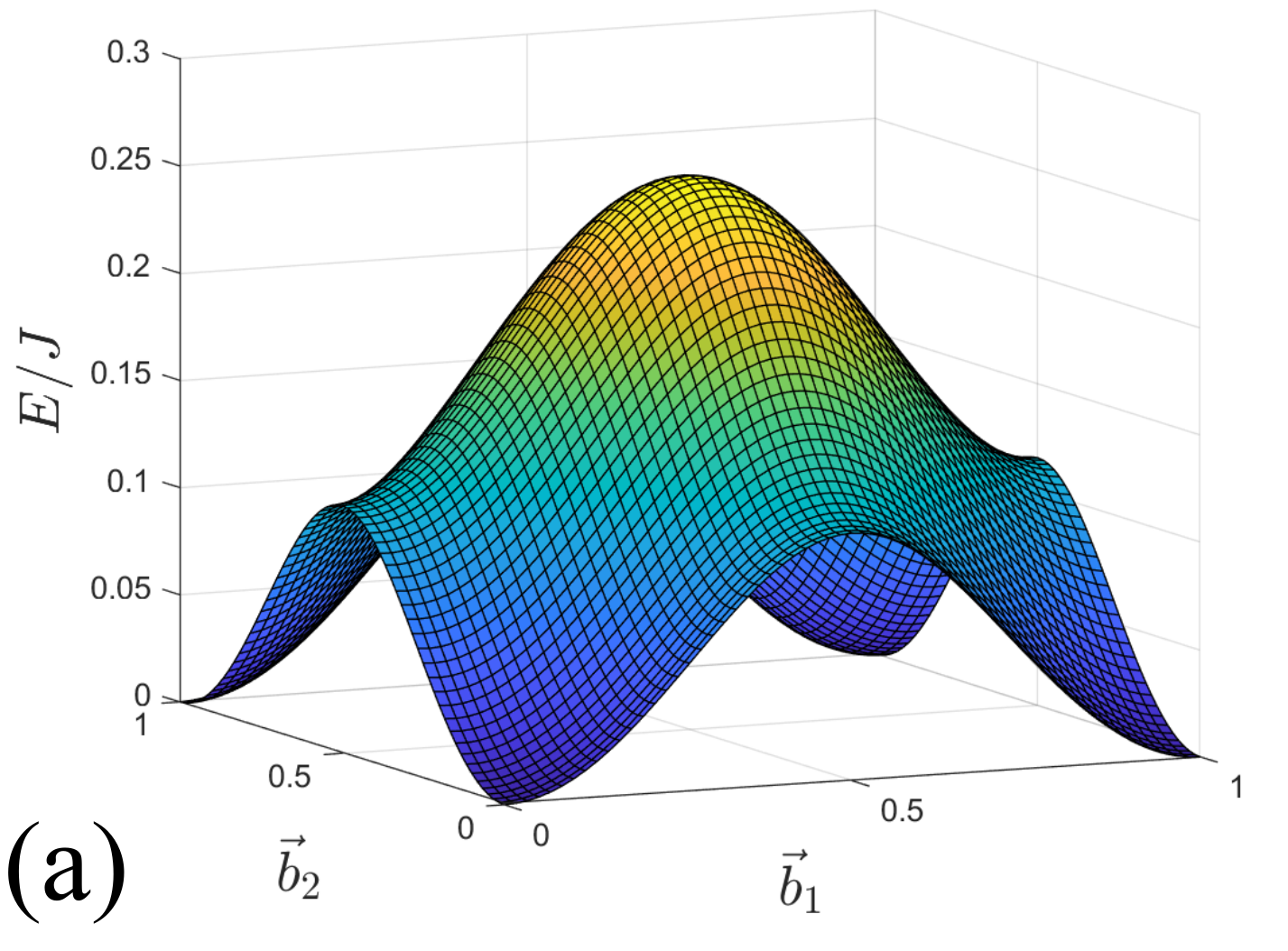}\!
\includegraphics[width=5.7cm]{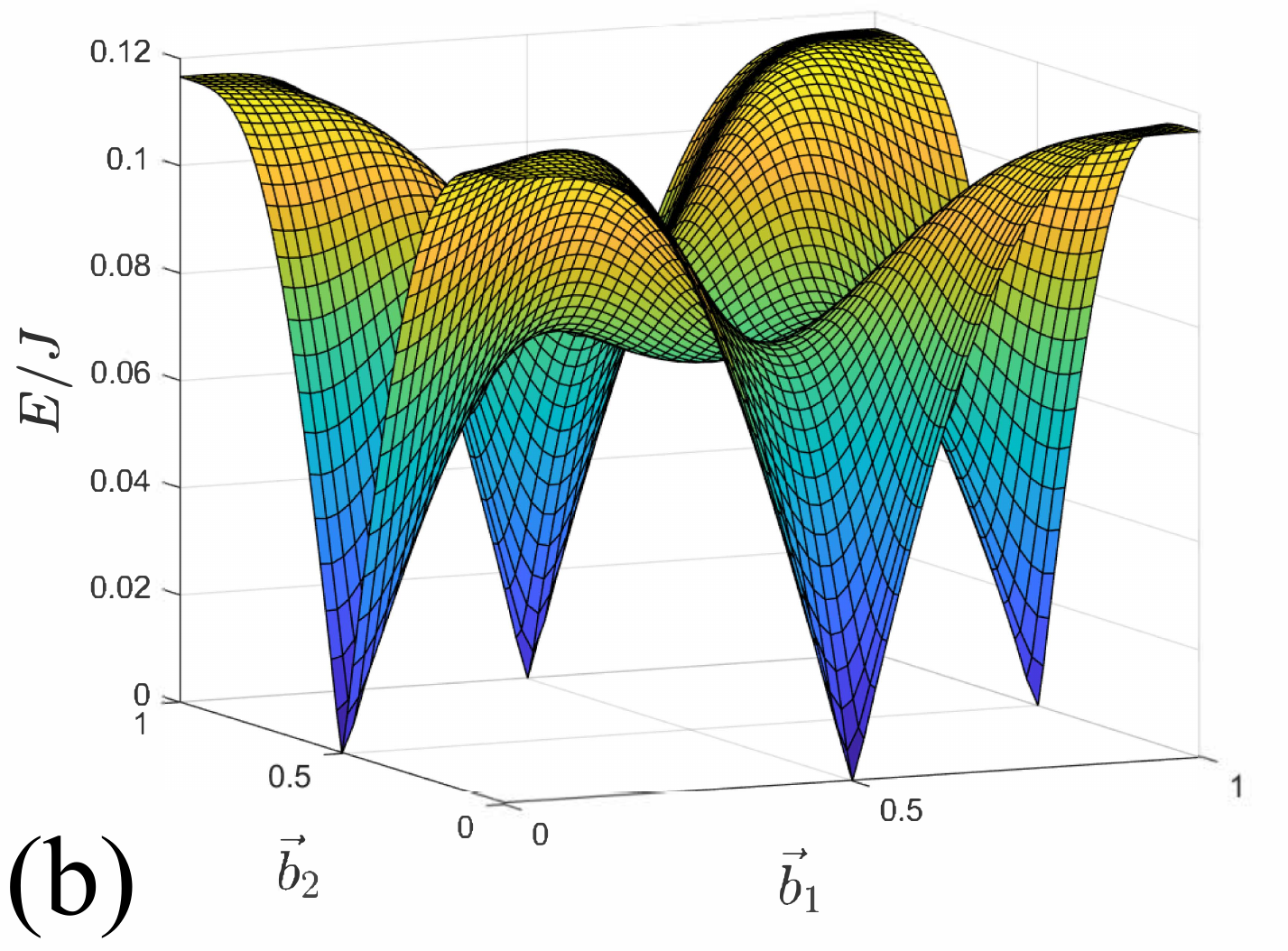}\!
\includegraphics[width=5.7cm]{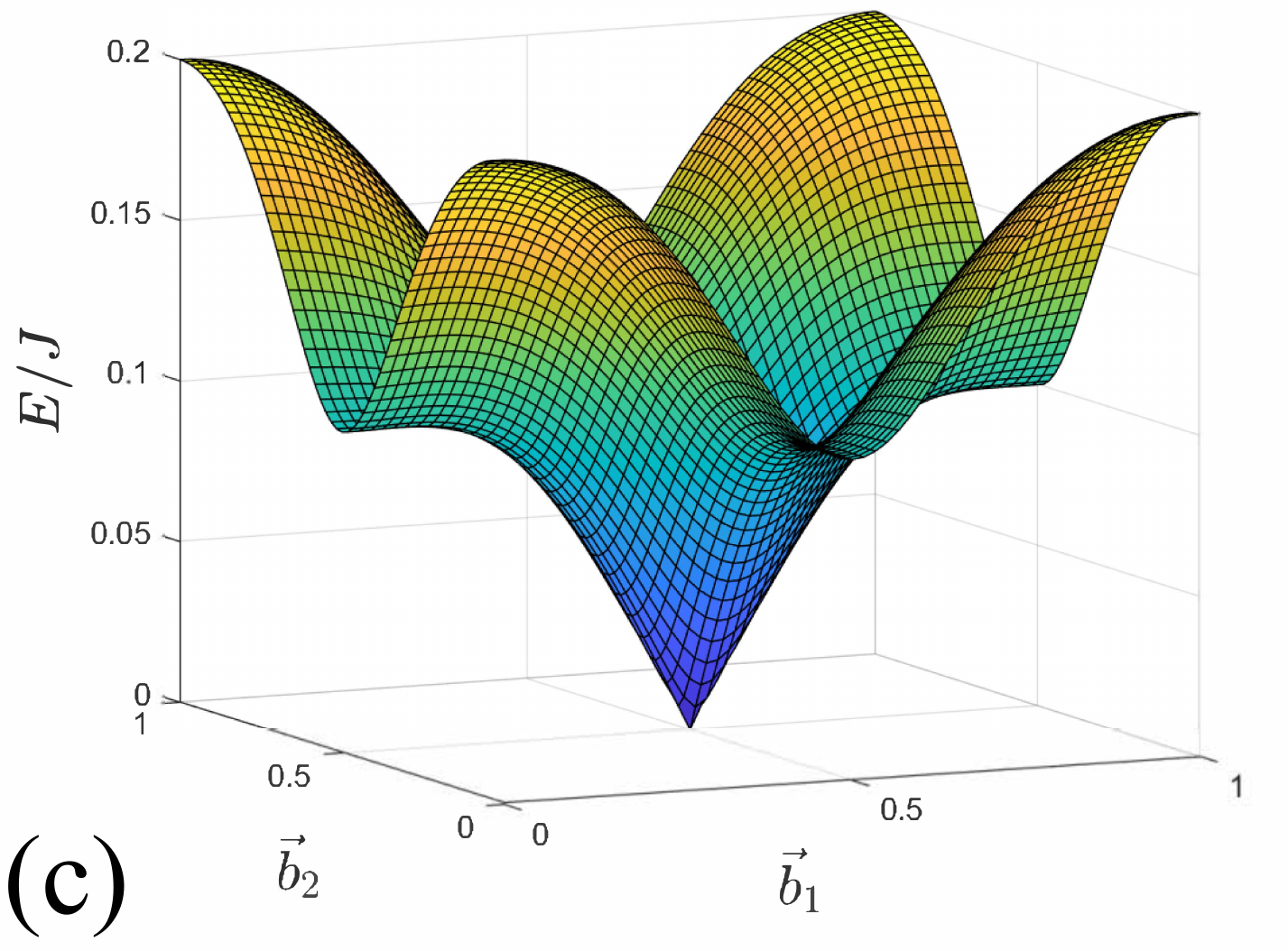}\!
\caption{Condensation of magnons (the lowest band in LSWT) at the upper critical point. (a) $\delta=0.8, \Gamma'/J=0.1$, the magnon gap closes at the $\pmb \Gamma$ point ($\pmb k=0$) with a quadratic dispersion; (b) $\delta$=0.85, $\Gamma^{\prime}/J$=0.2, the magnon gap closes at the $\pmb M_1$ point ($\pmb k=\pmb b_1/2$ or $\pmb k=\pmb b_2/2$) with a linear dispersion; (c) $\delta$=0.9, $\Gamma^{\prime}/J$=0.2 , the magnon gap closes at  point the $\pmb M_2$ point ($\pmb k=\pmb b_1/2 + \pmb b_2/2$) with a linear dispersion.
}\label{fig:SW}
\end{figure*}

\begin{figure*}[t]
\includegraphics[width=5.7cm]{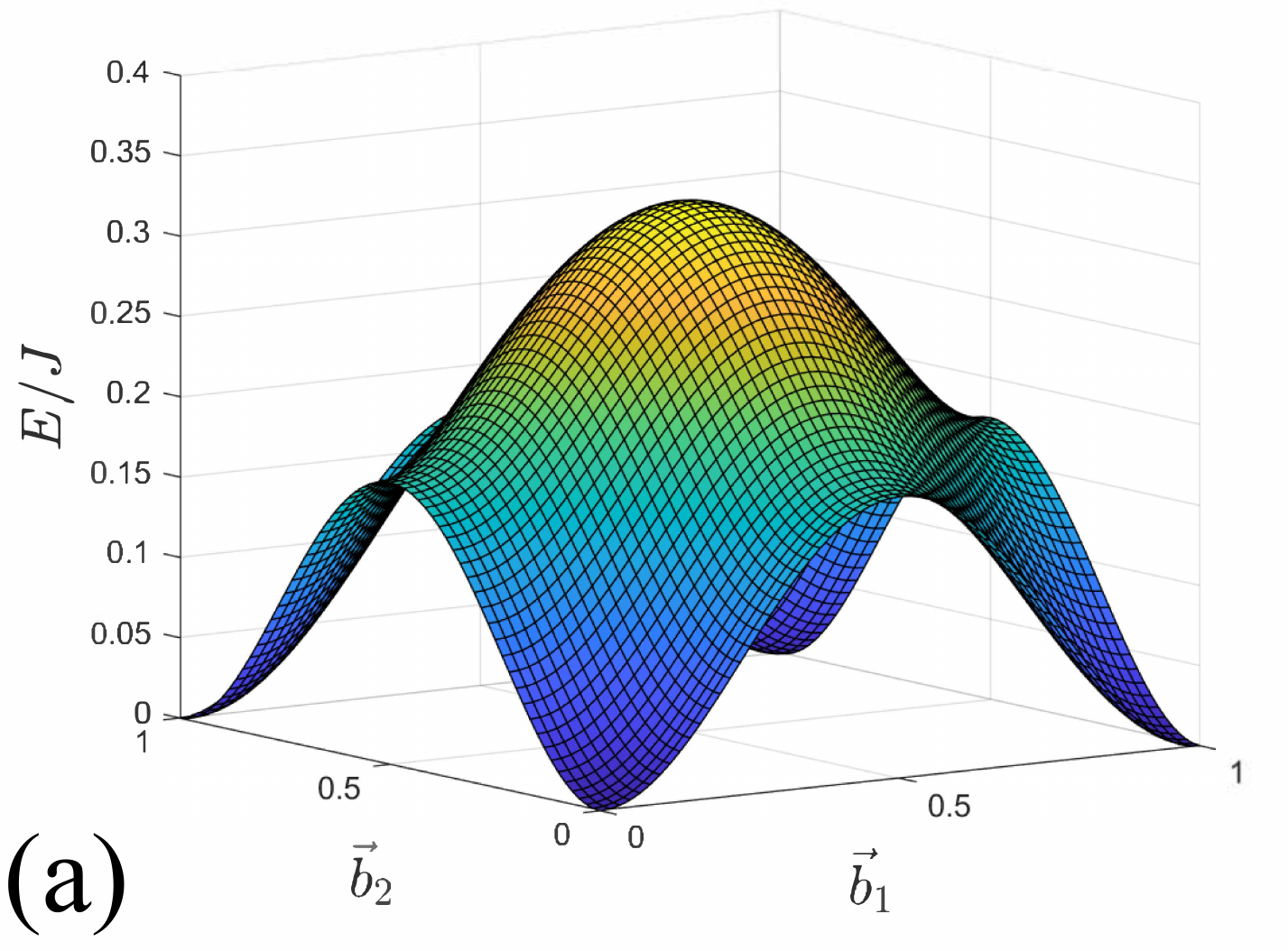}\!
\includegraphics[width=5.7cm]{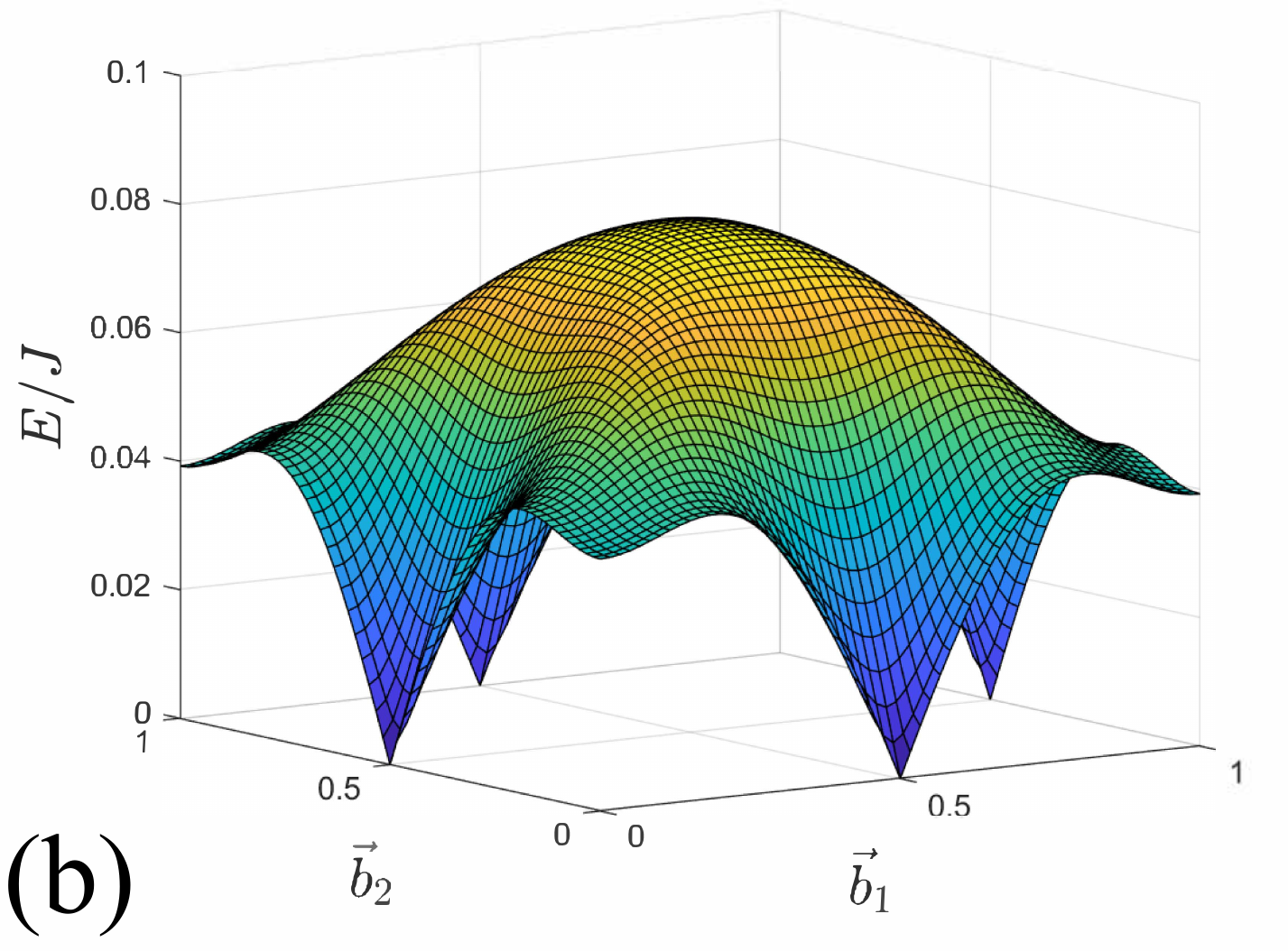}\!
\includegraphics[width=5.7cm]{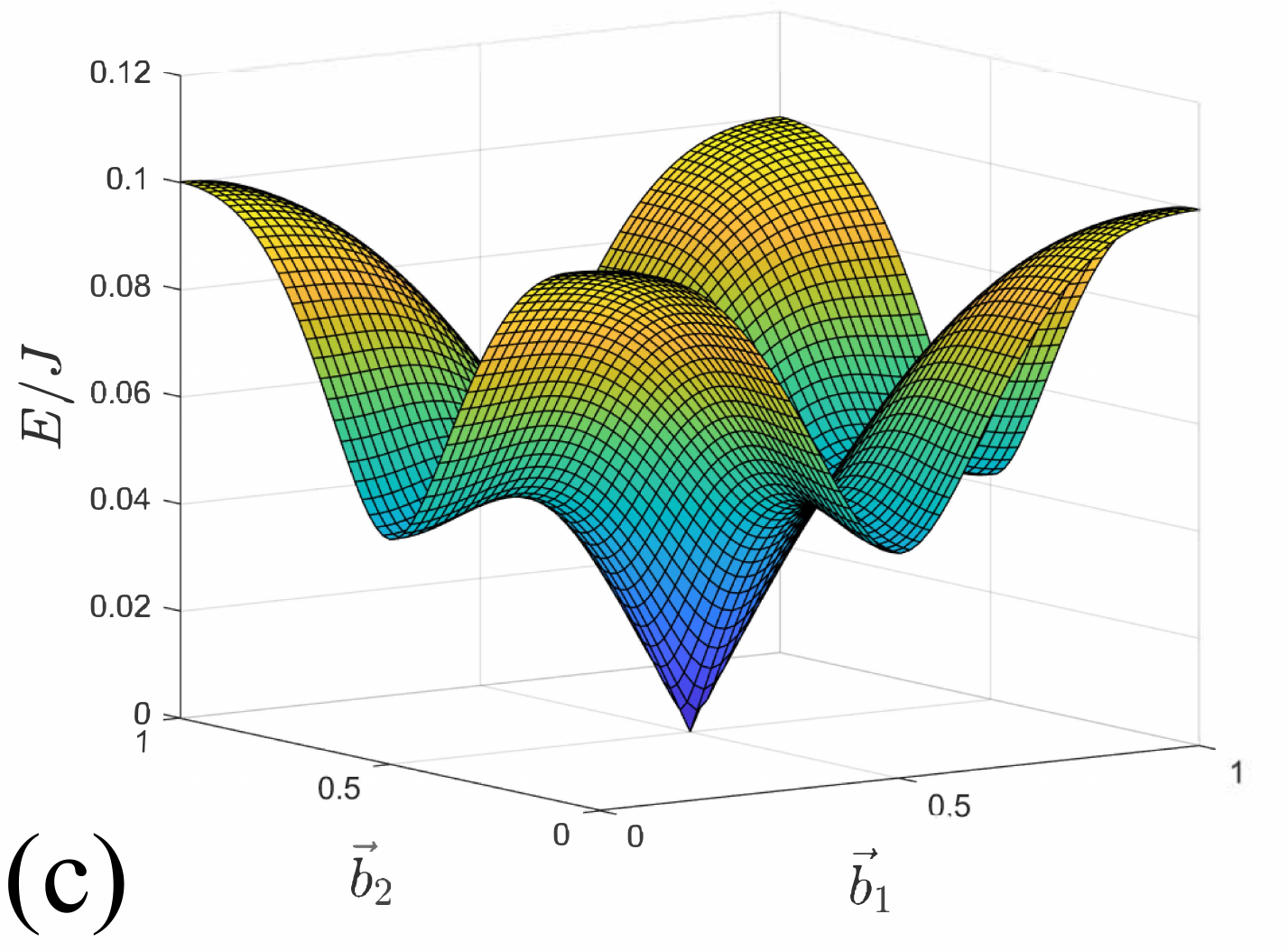}
\caption{Condensation of magnons (the lowest band in SU(4) LSWT) at the lower critical point.  (a) $\delta=0.8, \Gamma'/J=0.05$, the magnon gap closes at the $\pmb \Gamma$ point with a quadratic dispersion; (b) $\delta$=0.9, $\Gamma^{\prime}/J$=0.1, the magnon gap closes at the $\pmb M_1$ point with a linear dispersion; (c) $\delta$=0.95, $\Gamma^{\prime}/J$=0.1, the magnon gap closes at the $\pmb M_2$ point  with a linear dispersion.
}\label{fig:SU4}
\end{figure*}

Comparing Fig.\ref{fig:DSF_J1_0d9_Bz0d58}(a) and Fig.\ref{fig:DSF_J1_0d9_Bz0d58}(b), it can be seen that the lower magnon band contributes an enormous weight in the DSF data. In contrast, the second band in the renormalized excitations, which is dispersive, has a very small weight. However, there is indeed an almost flat band in the DSF data at around $\omega/J \approx 1,$ whose dispersion is qualitatively consistent with the upper magnon band in LSWT. Therefore, it is better not  to identify the second excitation band (the dashed line) in Fig.\ref{fig:DSF_J1_0d9_Bz0d58}(a) as the upper magnon band, instead, the upper magnon band locates somewhere inside the continuum. If this is true, then it means that there is a nonzero density of states between the lower and upper magnon bands, which is  quite unusual in quantum antiferromagnets. The weight of the DSF signal for the magnon bands almost vanishes at some momentum points, the reason might be that we only considered the $zz$-component. Finally,  the higher-energy continuum contributes small but nonzero weight to the DSF, showing that quantum fluctuations at $\omega > J$ (above the magnon bands) is not negligible.

The DSF for the intermediate stripe and zigzag ordered phases will not be shown here.

\section{Magnon condensation and critical properties}\label{SecV}
\subsection{Higher-field Critical Points}
In this section, we will adopt the Holstein–Primakoff (HP) representation\cite{HP} of spins and calculate the magnon spectrum. As illustrated above, at the critical point where the magnon gap closes, the magnons condense and the system undergoes a continuous phase transition. If the magnon gap closes at momentum $\pmb Q$ of BZ, then after the phase transition the system exhibits a long-range magnetic order whose static structure factor supports a peak at momentum $\pmb Q$.
To the end, we will find that spin-wave theory is qualitatively consistent with VMC calculations.

We will start with the polarized state in the high field limit, where the excitations are gapped bosonic magnons. Actually, when in the $B_c \to \infty$ limit, the system approximately recovers the $SO(2)$ symmetry (and hence the magnon-BEC picture approximately works). If the magnetic field is oriented along the $c$-direction, we can choose $c$-direction as the new $z$-axes. To this end, we perform a uniform orthogonal transformation ($\mathcal{R} \in SO(3)$) without changing the excitation spectrum,
\begin{equation}
\begin{aligned}
\begin{pmatrix}S_i^{\prime x} \\ S_i^{\prime y} \\ S_i^{\prime z}\end{pmatrix} =
\begin{pmatrix}\frac{1}{\sqrt{2}} & \frac{-1}{\sqrt{2}} & 0 \\ \frac{1}{\sqrt{6}} & \frac{1}{\sqrt{6}} & \frac{-2}{\sqrt{6}} \\ \frac{1}{\sqrt{3}} & \frac{1}{\sqrt{3}} & \frac{1}{\sqrt{3}} \end{pmatrix} \begin{pmatrix}S_i^{x} \\ S_i^{y} \\ S_i^{z}\end{pmatrix}.
\end{aligned}
\end{equation}
Thus the model (\ref{JGM}) is transformed into a new one under the above rotation matrix $\mathcal{R}$,
\beq\label{rotH}
H = \sum_{\langle i,j \rangle \in \gamma}\pmb S^{\prime\mathrm{T}}_i\mathcal{H}_{\gamma}^{\prime}\pmb S^{\prime}_j - \sum_{i}\pmb B^{\prime}\cdot\pmb{S}^{\prime}_i,
\eeq
where $\mathcal{H}_{\gamma}^{\prime}=\mathcal{R}\mathcal{H}_{\gamma}\mathcal{R}^\mathrm{T}$ and $\pmb B'= \pmb B \mathcal{R}^\mathrm{T}$.

To obtain the spin-wave spectrum, we will adopt the HP representation in the new spin frame on A-sublattice,
\beq\label{hp1}
S_i^{\prime x} &&=\frac{1}{2}\left(a_i^{\dag}\sqrt{1-a_i^{\dag}a_i}+\sqrt{1-a_i^{\dag}a_i}a_i\right)\sim{1\over2}(a_i^\dag+a_i),\nonumber \\
S_i^{\prime y} &&=\frac{i}{2}\left(a_i^{\dag}\sqrt{1-a_i^{\dag}a_i}-\sqrt{1-a_i^{\dag}a_i}a_i\right)\sim{i\over2}(a_i^\dag-a_i),\nonumber \\
S_i^{\prime z} &&={1\over2}-a_i^{\dag}a_i,
\eeq
where the bosons satisfy the usual commutation relations $[a_i, a_j^\dagger] = \delta_{ij}$.
We could construct HP representation on B-sublattice by the same way [$a_i \rightarrow b_i$ in (\ref{hp1})].
Substituting the above formulas into the rotated Hamiltonian (\ref{rotH}) and keeping the quadratic terms, we obtain the following Hamiltonian on the Fourier bases,
\beq\label{HSW}
H_{\rm SW} = \sum_{\pmb k} \Psi^\dagger(\pmb k) \mathcal{H}(\pmb k) \Psi(\pmb k),
\eeq
where $\Psi^\dagger(\pmb k)=(a_{\pmb k}^{\dag}, b_{\pmb k}^{\dag}, a_{\pmb {-k}}, b_{\pmb {-k}})$  with $a_{\pmb k}^\dag$ ($b_{\pmb k}^\dag$) being the magnon creation operator on sublattice A (B).
It is very important to note that, unlike the fermionic case, the eigenenergies and eigenmodes in this case are obtained by diagonalizing the non-Hermitian matrix $\Sigma\mathcal{H}(\pmb k)$, where $\Sigma={\rm diag}(1,1,-1,-1)$.
Thus diagonalizing the above non-Hermitian matrix, we get the magnon excitations spectrum (see Appendix \ref{app:highfield} for details).

The magnon gap decreases when lowering the field strength. At a critical field, the magnon gap closes at a certain momentum point in the BZ (see Fig.\ref{fig:SW}) and the system undergoes a continuous phase transition.  The momentum at which the magnon gap closes depends on the interaction parameters. If one further decreases the field strength, then the magnon spectrum in spin-wave Hamiltonian (\ref{HSW}) will have an imaginary part, indicating that the polarized phase is unstable. To avoid the instability in the excitation spectrum, the system enters an ordered phase where the spins exhibit a different configuration. In other words, the condensation of magnons results in a long-range magnetic order and the excitation spectrum changes accordingly.

If $\Gamma'$ term is small, the magnons condense at the $\pmb \Gamma$ point (i.e. $\pmb k=0$) and the system exhibits canted N\'{e}el order after the phase transition. As shown in Fig.\ref{fig:SW}(a), the dynamic critical exponent is $z=2$ at the transition point (owning to the quadratic dispersion), which is the same as the conventional magnon BEC\cite{Zapf2014} with $\Gamma'=0.$
On the other hand, if the magnons condense at the $\pmb M_1$ point (i.e. $\pmb k=\pmb b_1/2$ or $\pmb k=\pmb b_2/2$), then the system transits to the stripe phase and if the magnons condense at the $\pmb M_2$ point  (i.e. $\pmb k=\pmb b_1/2+\pmb b_2/2$), then the system exhibits zigzag order. In these cases, the dynamical exponent is $z=1$, as shown in Fig.\ref{fig:SW}(b)\&(c) where the magnon gap closes linearly. From the dynamic critical exponent, it is expected that the correlation length near the critical point is given by $\xi\sim |B_c-B_{\rm critical}|^{-\nu}$ with $\nu=1/z$ in two spatial dimension if the continuous $SO(2)$ symmetry is present\cite{Sachdev1994}.

The above conclusions are consistent with conventional Landau-Ginzberg theory. Effective actions can be constructed to interpret the response to a magnetic field from general symmetry principle and spin-wave calculations\cite{Ye2017,Ye2020,Ye2021} (see Appendix \ref{EFT} for details).
For example, when SOC is very weak, the effective action with the dynamic exponent $z=2$ describing the continuous phase transition from the high-field PP to the intermediate AFM$_{ab}$ phase is given by
\begin{align}
    \mathcal{S}
    =\int d\tau d^2r
    &\Big(\psi^*\partial_\tau\psi
    +v_{a_1}^2|\partial_{a_1}\psi|^2
    +v_{a_2}^2|\partial_{a_2}\psi|^2 \nonumber \\
    &-\mu |\psi|^2+U |\psi|^4 \Big),
\end{align}
where $ \mu=B_{\rm critical}-B_c$, $U>0$, and $\psi$ is a complex order parameter.
The universality class for the quantum phase transition (QPT) is nothing but the $z=2$ 3D XY universality class.
If SOC is strong enough, the effective action is switched to another one with dynamic exponent $z=1$ which describes the continuous phase transition from the high-field PP to the intermediate zigzag (or stripe) phase
\begin{align}
    \mathcal{S}
    =\int d\tau d^2r
    &\Big((\partial_\tau\psi)^2
    +v_{a_1}^2(\partial_{a_1}\psi)^2
    +v_{a_2}^2(\partial_{a_2}\psi)^2 \nonumber \\
    &-\mu \psi^2
    +U \psi^4 \Big),
\end{align}
where $\psi$ is a real order parameter.
And the universality class for the QPT may be the $z=1$ 3D Ising universality class.
It is important to note that these various effective actions produce all these quantum phases and corresponding excitation spectrum discovered by the microscopic calculations.
Therefore, SOC is crucial to understand the nature of the field-induced QCP beyond the conventional magnon BEC.

\begin{table}[t]
\centering
\begin{tabular}{c ||c |c |c }
\hline
\hline
Parameter           & Intermediate phase &  Lower QCP  & Upper QCP \\
\hline
$\Gamma'=0$         &  AFM$_{ab}$        & $z=2$     & $z=2$ \\
\hline
                    &  AFM$_{ab}$        & $z=2$     & $z=2$ \\
$\Gamma'\neq 0$     &  Stripe            & $z=1$     & $z=1$ \\
                    &  Zigzag            & $z=1$     & $z=1$ \\
\hline
\hline
\end{tabular}
\caption{The dynamic critical exponent of field-induced QCP with or without $\Gamma'$ interactions from spin-wave theory.}
\label{tab:QCP}
\end{table}

\subsection{Lower-field Critical Points}
Now we go to the low field limit. When SOC is weak, the system enters the dimerized phase when the magnetic field is weak.
As illustrated in the previous section, the magnon bands split into three branches (for details see Appendix \ref{app:1st})
and the gap of the lowest branch ($b_+$) decreases linearly with the field strength, the magnon gap closes and the system enters the intermediate ordered phase when the field strength reaches a critical point.

To obtain the critical exponent at the lower critical points, we consider each $z$-bond as an effective site, which contains four internal degrees of freedom. By regarding the four bases on each effective site as a `SU(4) spin',  namely, by mapping the four bases $|\!\!\up\up\rangle,  |\!\!\up\dn\rangle, |\!\!\dn\up\rangle,  |\!\!\dn\dn\rangle$ on the dimer bond into four species of  boson operators $$B^\dag= \big(b^{\dag}_{\up\up}, b^{\dag}_{\up\dn}, b^{\dag}_{\dn\up}, b^{\dag}_{\dn\dn}\big)$$
we can figure out the SU(4) spin wave\cite{LJXSU4}, where the `classical' ground state is a singlet formed by the two spin-1/2 spins on each effective site
\beq\label{eq:Dimer}
|{\rm Dimer}\rangle= \prod_i \Big[{1\over\sqrt2}(b_{\up\dn}^\dag-b_{\dn\up}^\dag)\Big]_i |{\rm vac}\rangle.
\eeq
By adopting the SU(4) HP approximation\cite{Muniz_2014} and keeping the quadratic terms only, we obtain the spin-wave Hamiltonian (see Appendix \ref{app:lowfield} for details of calculations), which can be diagonalized using bosonic Bogoliubov transformation.

The dispersions of the $b_+$ branch in SU(4) spin-wave theory are shown in Fig.\ref{fig:SU4}, it can be found that the dynamic critical exponent at the lower-field critical point is the same as that in the higher-field critical point (see Fig.\ref{fig:SW} for illustration).
Namely, the dynamic critical exponent is $z=2$ for the transition from the dimerized phase to the AFM$_{ab}$ phase and is $z=1$ for the transitions from the dimerized phase to the stripe phase or to the zigzag phase, which are summarized in Tab.\ref{tab:QCP}.
Meanwhile, based on symmetry analysis, the effective actions describing the continuous phase transition from low-field dimerized phase to intermediate ordered phases are similar to the ones from high-field PP to intermediate ordered phases and will not be repeated here.

When $\Gamma'$ interaction is large enough, the gap of $b_0$ branch will close, which means that the ground state will become N\'{e}el order where the spins are oriented perpendicular to the honeycomb plane (called AFM$_c$) through a continuous phase transition.
In this case, with the increasing of field strength, the system enters field-induced in-plane ordered phases by a first-order phase transition (see Appendix \ref{app:1st} for details).

\section{Conclusions and Discussions}\label{SecVI}
We studied the $J$-$\Gamma'$ model on the distorted honeycomb lattice in a magnetic field via variational Monte Carlo method. We conclude that in a dimerized antiferromagnet the external magnetic field will generally induce an intermediate phase with long-range AFM orders no matter if the ground state at zero field is a non-magnetic state with dimerized singlets or an AFM ordered state. The ordering pattern in the intermediate-field phase can be interpreted from the high-field or low-filed limit, where the magnon gap closes at a certain point in the Brillouin zone with decreasing or increasing of the magnetic field strength.
Starting from the general symmetry principle, we provide the effective field theory for these phase transitions.
While the phase transitions are continuous, the nature of the field-induced quantum critical points can be different from the conventional magnon BEC.

In obtaining the classical order, we have used the canted single-$\pmb Q$ approximation. We cannot rule out the possibility that the classical magnetic order is a multi-$\pmb Q$ state\cite{Janssen2016, YuanLi2021} instead of a single-$\pmb Q$ state. Therefore, the spin-orbit coupled spin model deserves to study with different numerical approaches.

 Although the experiments on the $4d$ or $4f$ dimerized magnetic materials such as $\alpha$-RuCl$_3$ and Yb$_2$Si$_2$O$_7$ motivate us to study the effect of SOC on magnon BEC, in the present work we are not aiming to interpret the experimental data for concrete materials but to address a general question: what is the possible consequence of SOC on dimerized antiferromagnets. Notably, to interpret the experimental observations in $\alpha$-RuCl$_3$ and Yb$_2$Si$_2$O$_7$, one may need  to consider more interactions, and carefully tune the parameters.  We will leave this for future study.

\section*{Acknowledgements}
We specially thank Jinwu Ye for suggesting us to construct effective field theory on the quantum phase transitions and for very helpful discussions on technical issues. J.W. thanks Bo Li and Chuan Chen for valuable discussions and comments.
This work is supported by the NSF of China (No. 11574392), the National Natural Science Foundation of China (No. 12134020), the Fundamental Research Funds for the Central Universities and the Research Funds of Renmin University of China (No. 19XNLG11), and the China Postdoctoral Science Foundation (No. 2021M690093).

\appendix

\section{Phase Transitions To and From AFM$_c$} \label{app:1st}

\subsection{Relation between magnon condensation and in-plane AFM order} 

The moment at which the magnon closes gap and condenses determines the pattern of the consequent magnetic order. Suppose that the magnons condense at momentum $\pmb k$, then Bogoliubov quasi-particle $\alpha_{\pmb k}$ obtain non-zero expectation values in the ground state. Noticing that the HP bosons $a, a^\dag$ are nothing but spin operators $S^+, S^-$ (according to the polarized state), so $\langle S^{\tilde x}_i+iS^{\tilde y}_i\rangle \propto \langle \alpha_{\pmb k} \rangle e^{-i\pmb k\cdot \pmb r_i}$
have nonzero expectation values (notice that here $\tilde x$  and $\tilde y$ stand for ‘in-plane’ directions perpendicular to the c-axes). Therefore, the gap closing at $\pmb \Gamma$ or $\pmb M_1$ or $\pmb M_2$ point (here we have folded all the $\pmb k$ points to the first BZ) amounts to AFM$_{ab}$, stripe or zigzag order, respectively. Note that the relation between the momentum of the condensed boson and the pattern of the resulting in-plane orders is discussed more accurately in Appendix \ref{EFT}.

On the other hand, in the SU(4) spin-wave theory, if the ${1\over\sqrt2}(b_{\uparrow\downarrow} + b_{\downarrow\uparrow})$ boson condenses in the dimer state (eigenstate of $b_{\uparrow\downarrow} - b_{\downarrow\uparrow} $), then the ground state exhibits nonzero expectation values of $\langle b_{\uparrow\downarrow}\pm b_{\downarrow\uparrow} \rangle$ for each pair of spins on the strong bonds (here $\uparrow, \downarrow$ are eigenstates of $S^c=\pmb S \cdot \pmb c$). This indicates that the ground state exhibits nonzero AFM correlation for the $S^c$ operators, namely, the system enters the AFM$_c$ phase.

\subsection{From the dimerized phase to AFM$_c$ order}\label{app:1stdim}

\begin{figure}[t]
\includegraphics[width=7.5cm]{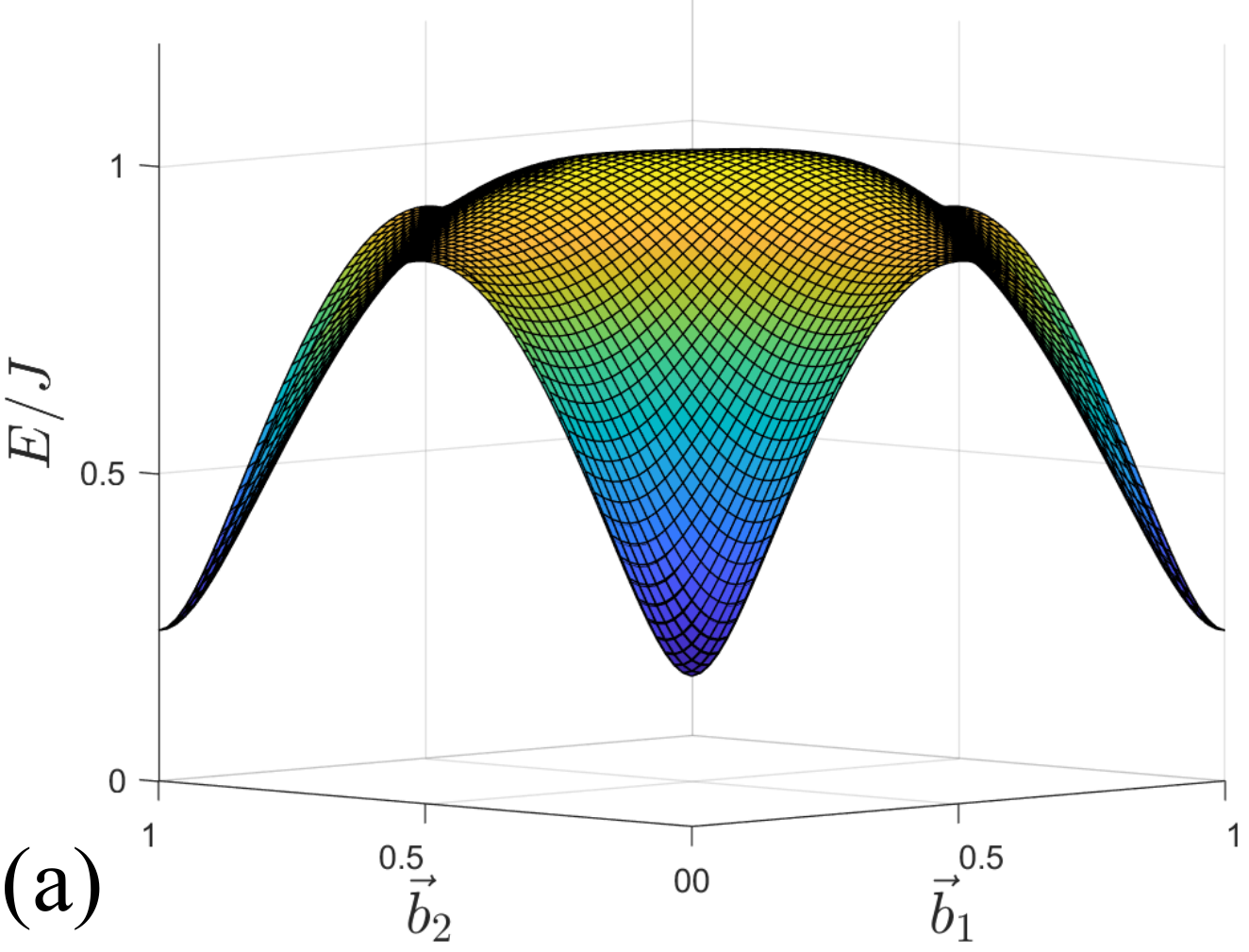}
\includegraphics[width=7.5cm]{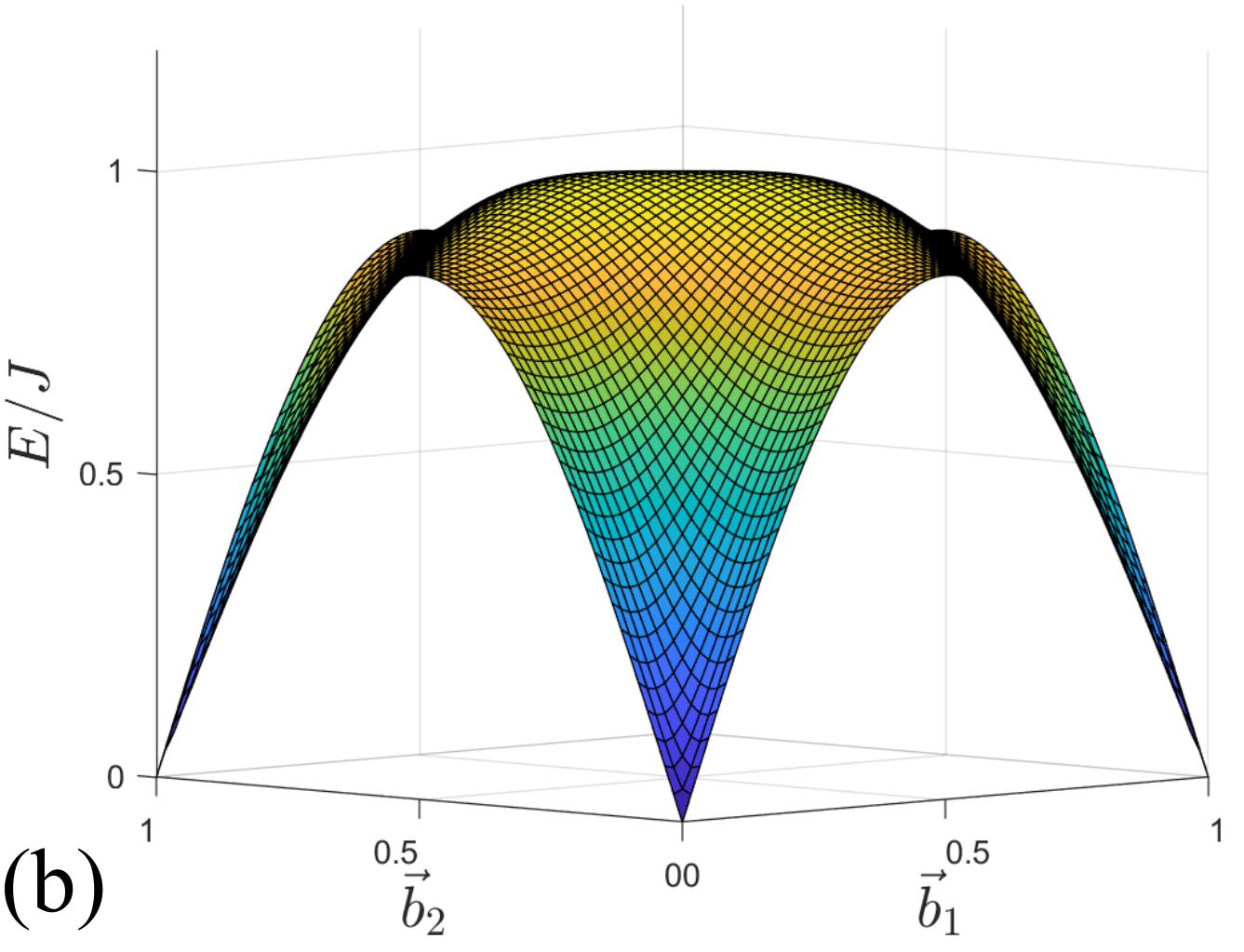}
\caption{
The spin-wave spectrum of AFM$_c$ order. (a) $\delta=0.5$, $\Gamma'/J=0.01$, $B_c/J=0$. The AFM$_c$ order appears here and it becomes a gapped state due to nonzero $\Gamma'$ interactions. (b) $\delta=0.5$, $\Gamma'/J=0$, $B_c/J=0$. AFM order is gapless with a linear dispersion around $\pmb \Gamma$-point (owning to continuous symmetry breaking).
}\label{fig:SW_AFMc}
\end{figure}

\begin{figure*}[t]
\includegraphics[width=5.7cm]{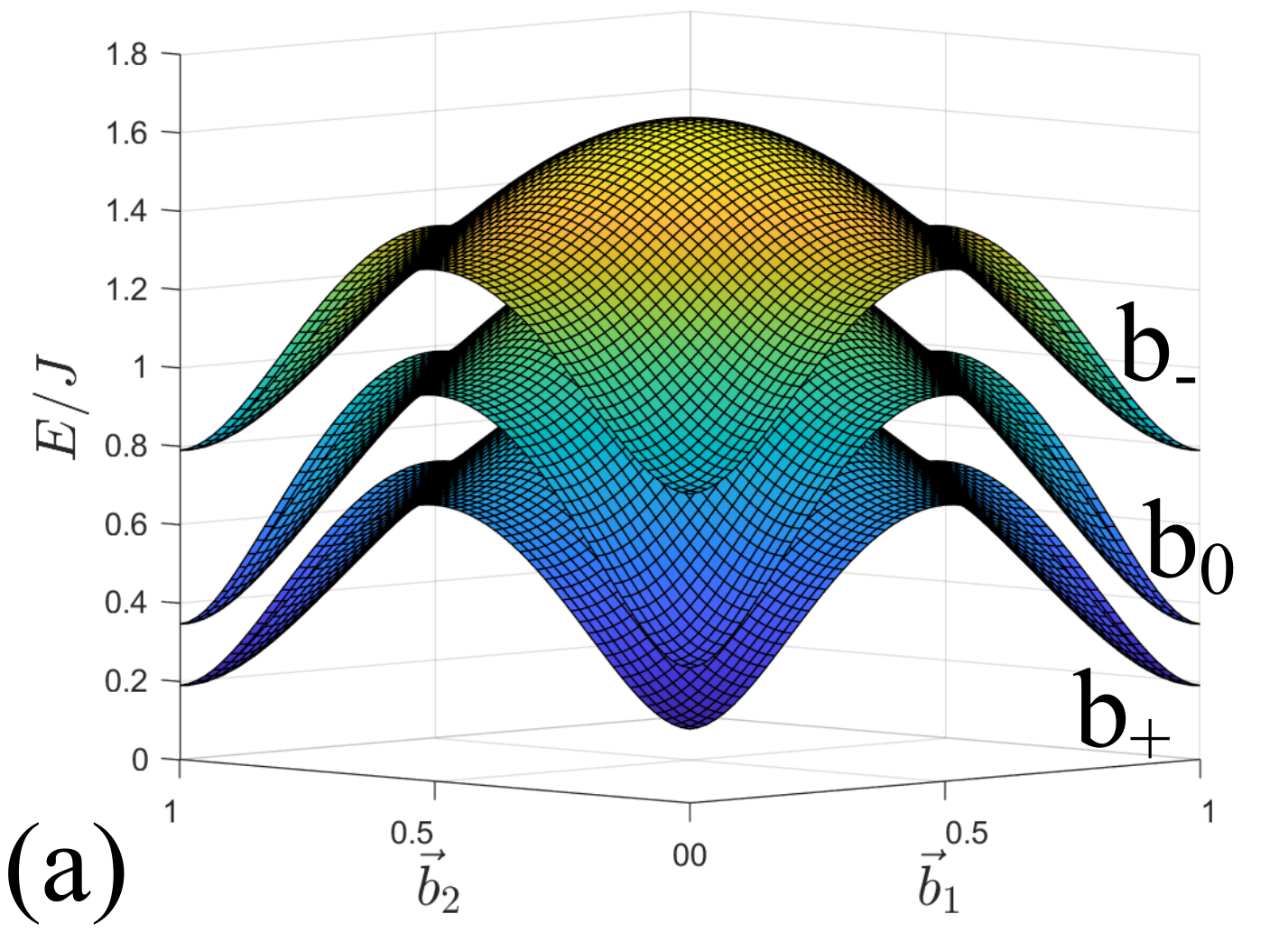}\!
\includegraphics[width=5.7cm]{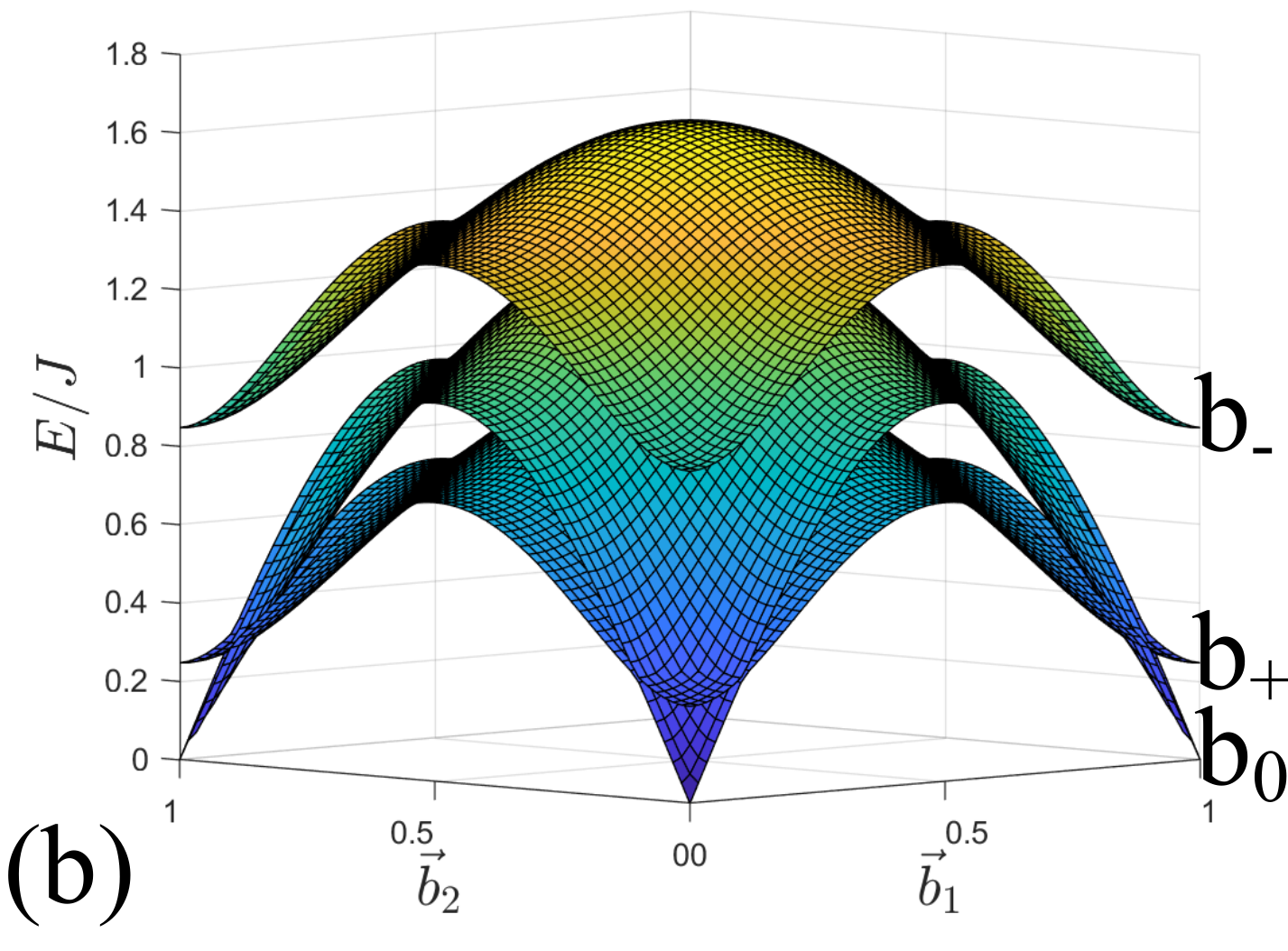}\!
\includegraphics[width=5.7cm]{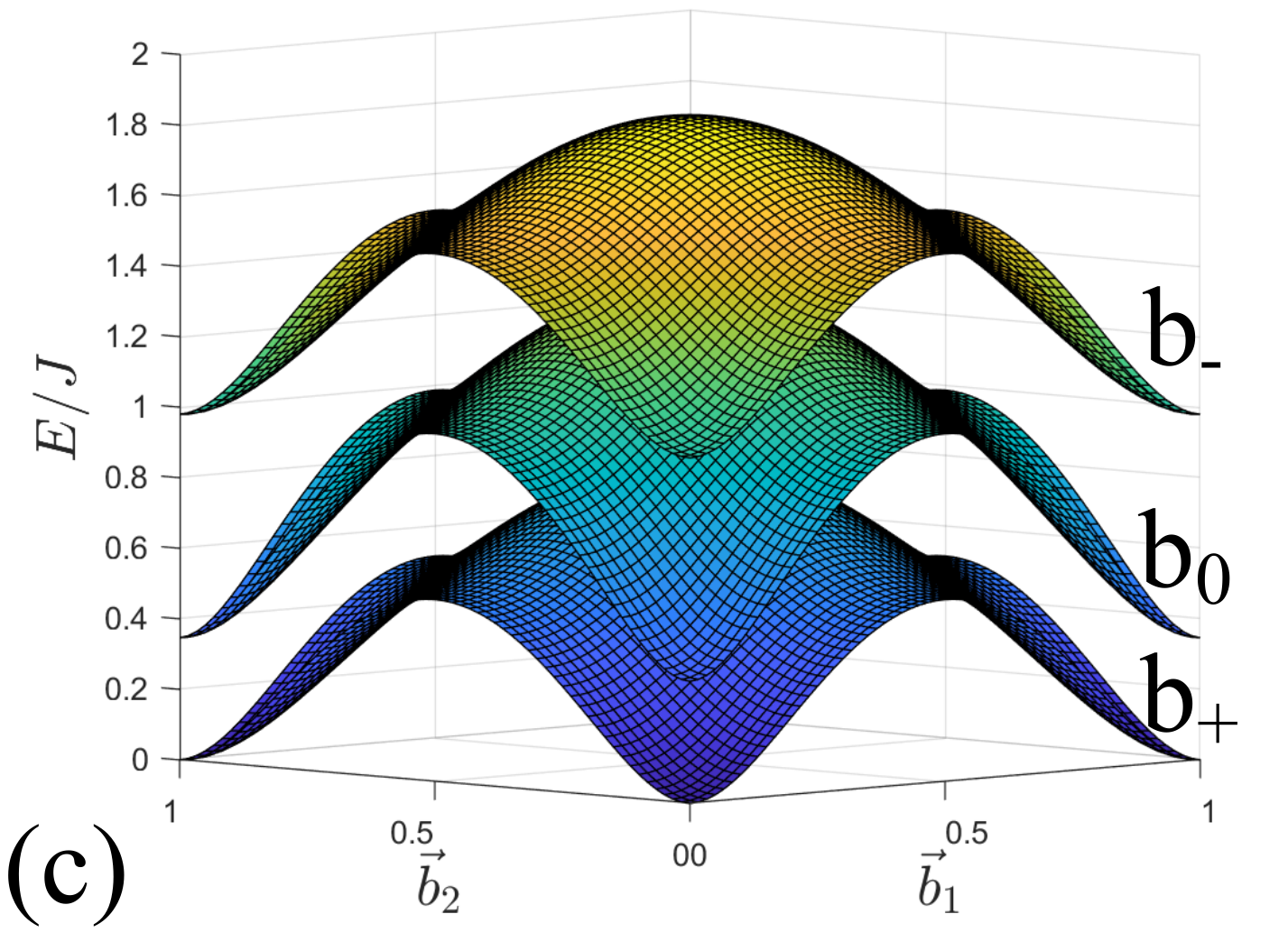}
\caption{
The SU(4) spin-wave spectrum. (a) $\delta=0.6$, $\Gamma'/J=0.02, B_c/J=0.3$. The dimerized ground state appears here and three magnon bands (namely, $b_+$, $b_0$, and $b_-$, respectively) split from each other due to nozero magnetic fileds. (b) $\delta=0.6$, $\Gamma'/J=0.05, B_c/J=0.3$. The gap of $b_0$-band closes, and the condensation of $b_0$-band indicates the continuous transition from the dimerized state to AFM$_c$ with a N\'{e}el pattern in the $c$-direction. (c) $\delta=0.6$, $\Gamma'/J=0.02, B_c/J=0.49$. The gap of $b_+$-band closes, and the condensation of $b_+$-band indicates a continuous transition from the dimerized state to AFM$_{ab}$ with canted axis along $c$-direction.
}\label{fig:SU4GpBc}
\end{figure*}

When the magnetic field $B_c=0$, with the increasing of $\Gamma'$ interactions, the system undergoes a continuous phase transition from the dimerized phase to the AFM$_c$ phase where the spin orientation is parallel to the $c$-direction (see Fig.\ref{fig:PD01}).
It's important to note that mganon excitations of AFM$_c$ (except $\Gamma'$=0, $\delta \leq0.5$ line) is gapped from linear spin-wave theory as shown in Fig.\ref{fig:SW_AFMc}.
A natural question is how to understand the physical mechanism of phase transition.
Because the dimerized state has no classical counterpart unlike AFM order, it's difficult to address this question by conventional LSWT from AFM$_c$ order to dimerized phase.
Hence, we will address this question by SU(4) spin-wave theory (see Appendix \ref{app:lowfield}) from dimerized phase to AFM$_c$ order.
In the dimerized phase, three magnons bands split from each other due to the appearance of magnetic fields.
We call the middle magnon band $b_0$, the lowest band $b_+$, and highest band $b_-$, respectively [see Fig.\ref{fig:SU4GpBc}(a)].
Meanwhile, the gap of $b_0$-band is independent on the $c$-direction field while the gap of $b_+$-band or $b_-$-band is dependent on the field.
In other words, the gap of $b_0$-band does not respond to magnetic fields, which means the $b_0$-band supports an antiferromagnetic pattern in the direction of the magnetic field.
Thus the mechanism behind the phase transition from dimerized phase to AFM$_c$ phase is that the gap of $b_0$-band (in SU(4) spin-wave theory) will close as $\Gamma'$ interactions increase.
This condensation is at $\pmb \Gamma$ point with dynamic critical exponent $z=1$ [see Fig.\ref{fig:SU4GpBc}(b)].

\subsection{From AFM$_c$ to the intermediate ordered phases}
\begin{figure*}[t]
\includegraphics[width=5.7cm]{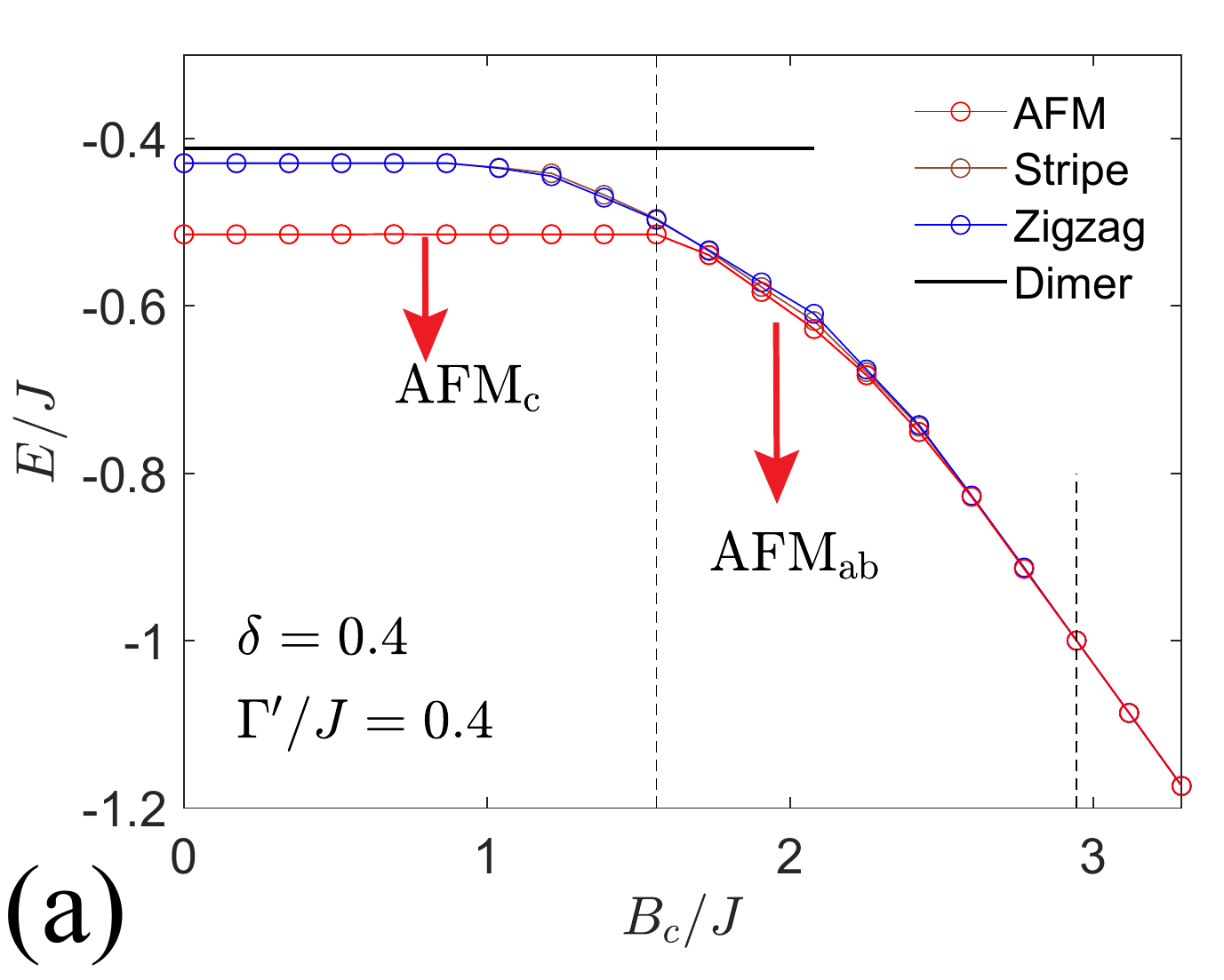}\!
\includegraphics[width=5.7cm]{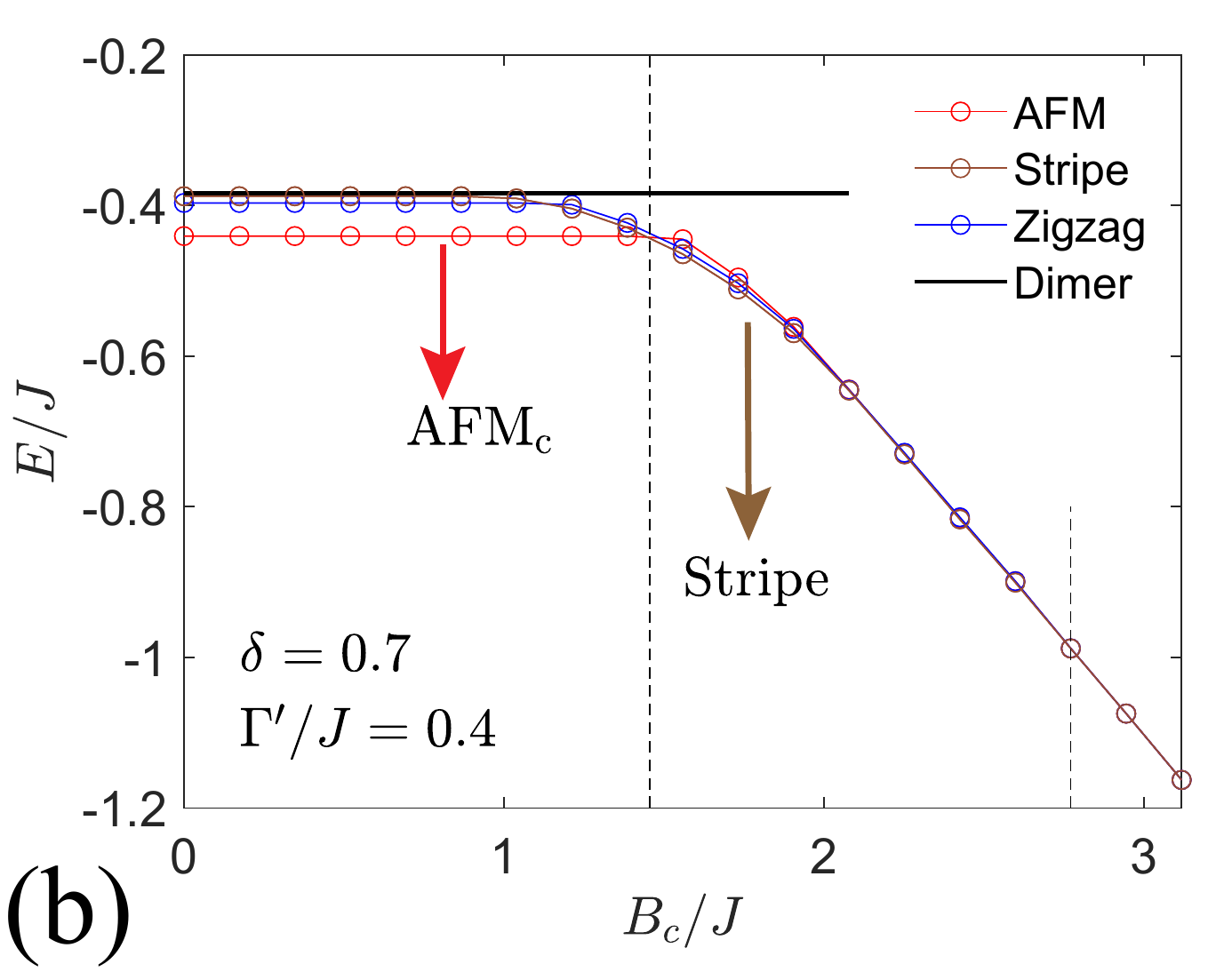}\!
\includegraphics[width=5.7cm]{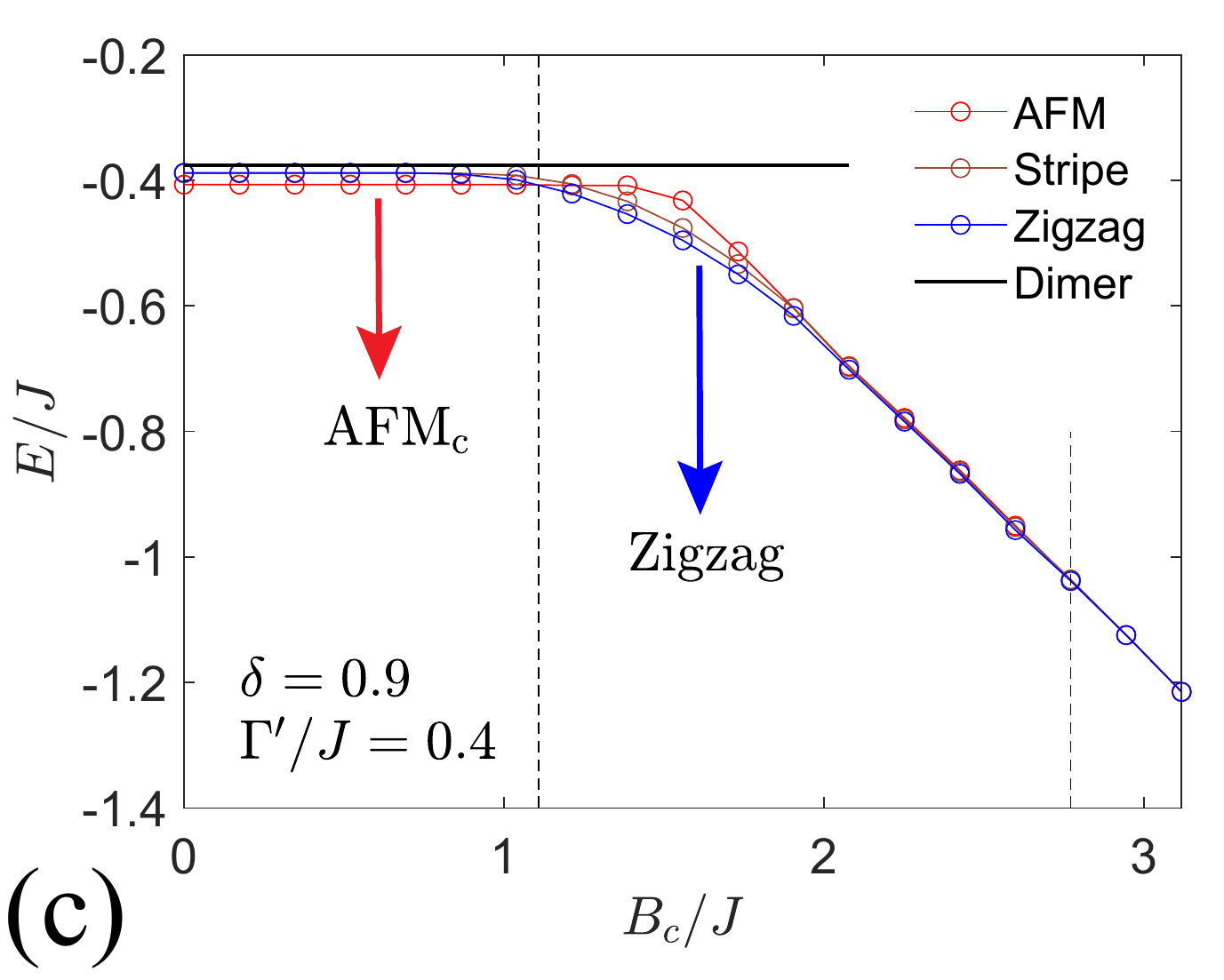}\!
\caption{Energy curves of the ground state with the increasing magnetic field (in the parameter region of AFM$_c$ at zero field). (a) $\delta$=0.4, $\Gamma^{\prime}/J$=0.4, (b) $\delta$=0.7, $\Gamma^{\prime}/J$=0.4, (c) $\delta$=0.9, $\Gamma^{\prime}/J$=0.4 are points in region-IV, region-V, and region-VI with circle [see Fig.\ref{fig:PD01}(b)], respectively. The vertical dashed lines indicate the values of the critical fields.
}\label{fig:levelcrossing}
\end{figure*}

When turning on the magnetic field $B_c$, the AFM$_{\rm c}$ state will undergo two successive phase transitions from VMC calculations.
The first transition is of first order, and the second one is of second order. The first-order phase transition is reflected by the level crossing of the energy curve of the ground state, as shown in Fig.\ref{fig:levelcrossing}.
It is important to note that the energy difference between the ground state and other ordered states in the intermediate field region is very visible (at least $10^{-2}J$ per site) although the difference between them looks small in Fig.\ref{fig:levelcrossing}.


The occurrence of the first-order phase transition can also be understood by comparing the nature of the two phases from the SU(4) spin-wave theory (see Appendix \ref{app:lowfield}). To be specific, the in-plane AFM order (including the AFM$_{ab}$, the stripe and the zigzag phases) results from the condensation of the $b_+$-band (at $\pmb \Gamma$, $\pmb M_1$ or $\pmb M_2$ point, respectively), see Fig.\ref{fig:SU4GpBc}(c) for a typical example. Noticing that for any given parameters $\Gamma'$ and $\delta$, there is {\it only one} intermediate phase between the low-field region and the high-field polarized region. Therefore, the nature of the intermediate in-plane AFM ordered phases in the whole phase diagram are essentially the same given that the ordering pattern is the same. On the other hand, the AFM$_c$ order can be attribute to the condensation of the  $b_0$-band [at the $\pmb \Gamma$ point, see Fig.\ref{fig:SU4GpBc}(b)]. Hence, the phase transition between the AFM$_c$ phase and the in-plane AFM phases should be of first order.

Depending on the ordering pattern of the intermediate phases (namely, AFM$_{\rm ab}$, stripe or zigzag), the AFM$_{\rm c}$ phase splits into three regions labeled as IV, V, VI, respectively.
It should be noted that the gapless line with $\Gamma'=0, \delta<0.5$ is special, where the transition to the AFM$_{ab}$ phase is of second order and the critical field strength is zero. Namely, on this line an infinitesimally small magnetic field can drive the system to the AFM$_{ab}$ phase.

Finally, we note that it is difficult to obtain the first-order phase transition purely from LSWT (with AFM$_c$ order as a starting point).
From the magnon dispersion of the AFM$_c$ state, one can still observe the gap closing when increasing the field. Consequently, it seems that the system will undergo a second-order phase transition into another intermediate ordered phase, e.g. an AFM$'_{ab}$ phase. But this AFM$'_{ab}$ phase has finite AFM order along the $c$-direction, so it differs greatly from the previously mentioned AFM$_{ab}$ phase which has uniform polarization along the $c$-direction. Our VMC calculations show that before the magnon gap closing the system has already entered the AFM$_{ab}$ phase via a first order phase transition. In other words, with increasing of the magnetic field, the AFM$_c$ order vanishes with a jump and the AFM$'_{ab}$ phase never occurs in the phase diagram. The above discussion also applies for the first-order transition from the AFM$_c$ phase to the intermediate stripe phase or the zigzag phase.

\section{Spin-Wave Spectrum at High Fields}\label{app:highfield}
To study the condensation of bosons from the polarized phase, we adopt the linear spin-wave theory.
Firstly, constructing rotation matrix $\mathcal{R}$ such that the z-direction of the local coordinate system is parallel to the spin orientation.
\beq
\pmb{S^{\prime}}= \begin{pmatrix}S^{\prime x} \\ S^{\prime y} \\ S^{\prime z}\end{pmatrix} = \mathcal{R}\pmb{S} =
\begin{pmatrix}\frac{1}{\sqrt{2}} & \frac{-1}{\sqrt{2}} & 0 \\ \frac{1}{\sqrt{6}} & \frac{1}{\sqrt{6}} & \frac{-2}{\sqrt{6}} \\ \frac{1}{\sqrt{3}} & \frac{1}{\sqrt{3}} & \frac{1}{\sqrt{3}} \end{pmatrix} \begin{pmatrix}S^x \\ S^y \\ S^z\end{pmatrix}
\eeq
hence the $S^{\prime z}$ is along the original $c$-direction.

Secondly, we rewrite the original Hamiltonian as
\beq
H = \sum_{\langle i,j \rangle \in \gamma}\pmb S^\mathrm{T}_i \mathcal{H}_{\gamma}\pmb S_j - \sum_{i}\pmb B\cdot\pmb{S}_i,
\eeq
where
\begin{small}
\beq
\mathcal{H}_x = \begin{pmatrix} J_x & \Gamma' & \Gamma' \\ \Gamma' & J_x & 0 \\ \Gamma' & 0 & J_x \end{pmatrix},
\mathcal{H}_y = \begin{pmatrix} J_y & \Gamma' & 0 \\ \Gamma' & J_y & \Gamma' \\ 0 & \Gamma' & J_y \end{pmatrix},
\mathcal{H}_z = \begin{pmatrix} J_z & 0 & \Gamma' \\ 0 & J_z & \Gamma' \\ \Gamma' & \Gamma' & J_z \end{pmatrix}\nonumber
\eeq
\end{small}

Then we obtain the rotated Hamiltonian that is given by
\beq\label{RJGM}
H &=& \sum_{\langle i,j \rangle \in \gamma}\pmb S^\mathrm{T}_i\mathcal{R}^\mathrm{T}\mathcal{R}\mathcal{H}_{\gamma}\mathcal{R}^\mathrm{T}\mathcal{R}\pmb S_j- \sum_{i}\pmb B\mathcal{R}^\mathrm{T}\cdot\mathcal{R}\pmb{S}_i \nonumber\\
 &=& \sum_{\langle i,j \rangle \in \gamma}\pmb S^{\prime\mathrm{T}}_i\mathcal{H}_{\gamma}^{\prime}\pmb S^{\prime}_j - \sum_{i}\pmb B^{\prime}\cdot\pmb{S}^{\prime}_i
\eeq
and $\pmb{B}^{\prime}$ is in the same direction with $S^{\prime z}$.
For convenience of latter discussion, we introduce the notation
\beq\label{entry}
\mathcal{H}_{\gamma}^{\prime}=\begin{pmatrix} h^{\prime 11}_{\gamma} & h^{\prime 12}_{\gamma} & h^{\prime 13}_{\gamma} \\ h^{\prime 21}_{\gamma}& h^{\prime 22}_{\gamma}& h^{\prime 23}_{\gamma} \\ h^{\prime 31}_{\gamma}& h^{\prime 32}_{\gamma} & h^{\prime 33}_{\gamma}\end{pmatrix}
\eeq
In the rotational basis, we perform a HP expansion on A-sublattice
\beq\label{hp}
S_i^{\prime x} &&=\frac{1}{2}\left(a_i^{\dag}\sqrt{1-a_i^{\dag}a_i}+\sqrt{1-a_i^{\dag}a_i}a_i\right)\sim{1\over2}(a_i^\dag+a_i),\nonumber \\
S_i^{\prime y} &&=\frac{i}{2}\left(a_i^{\dag}\sqrt{1-a_i^{\dag}a_i}-\sqrt{1-a_i^{\dag}a_i}a_i\right)\sim{i\over2}(a_i^\dag-a_i),\nonumber \\
S_i^{\prime z} &&={1\over2}-a_i^{\dag}a_i,
\eeq
and the expansion on B-sublattice is similar.
Keeping only terms that contribute up to quadratic order in the Hamiltonian,
we will obtain the following spin-wave Hamiltonian in the Fourier bases,
\beq
H_{\rm SW} = \sum_{\pmb k} \Psi^\dagger(\pmb k) \mathcal{H}(\pmb k) \Psi(\pmb k),
\eeq
where $\Psi^\dagger(\pmb k)=(a_{\pmb k}^{\dag}, b_{\pmb k}^{\dag}, a_{\pmb {-k}}, b_{\pmb {-k}})$  with $a^\dag$ ($b^\dag$) being the magnon creation operator on sublattice A (B).
And $\mathcal{H}(\pmb k)$ is $4\times 4$ matrix of the form\cite{Kim2021}
\begin{align}
\mathcal{H}(\pmb k) = \frac{1}{2} \begin{pmatrix} \mathcal{A}_{\pmb k} & \mathcal{B}_{\pmb k} \\ \mathcal{B}_{\pmb {-k}}^* & \mathcal{A}_{\pmb {-k}}^\mathsf{T} \end{pmatrix}.
\end{align}
with $\mathcal{A}_{\pmb k}$ and $\mathcal{B}_{\pmb k}$ being 2 dimensional matrices.
The forms are as follows,
\begin{align}
&\mathcal{A}_{\pmb k}=\begin{pmatrix} B' & 0 \\ 0 & B' \end{pmatrix} + \sum_\gamma \begin{pmatrix} -h^{\prime 33}_{\gamma} & \frac{h^{\prime 11}_{\gamma}+h^{\prime 22}_{\gamma} }{2}e^{i\pmb k\cdot \delta_\gamma}  \\ \frac{h^{\prime 11}_{\gamma}+h^{\prime 22}_{\gamma} }{2}e^{-i\pmb k\cdot \delta_\gamma} & -h^{\prime 33}_{\gamma} \end{pmatrix}, \nonumber \\
&\mathcal{B}_{\pmb k} = \sum_\gamma \frac{1}{2}(h^{\prime 11}_{\gamma}+2i h^{\prime 12}_{\gamma}-h^{\prime 22}_{\gamma}) \begin{pmatrix} 0 & e^{i\pmb k\cdot \delta_\gamma} \\ e^{-i\pmb k\cdot \delta_\gamma} & 0 \end{pmatrix}, \nonumber
\end{align}
where $\delta_\gamma$ indicates the distance between the the nearest-neighbor sites on $\gamma$-bond.
Note that only entries $h^{\prime 11}_{\gamma}$, $h^{\prime 12}_{\gamma}$, $h^{\prime 21}_{\gamma}$, $h^{\prime 22}_{\gamma}$, and $h^{\prime 33}_{\gamma}$ in (\ref{entry}) are relevant to the linear spin-wave Hamiltonian above.
The eigenenergies in this case are obtained by diagonalizing the matrix $\Sigma\mathcal{H}(\pmb k)$, where $\Sigma={\rm diag}(1,1,-1,-1)$.
The bosonic Bogoliubov transformation matrice $\mathcal{U}$ satisfies the condition $\mathcal{U}^\dag \Sigma \mathcal{U}=\mathcal{U} \Sigma \mathcal{U}^\dag = \Sigma$. We will describe the process to obtain the bosonic Bogoliubov transformation\cite{Lake2015} in the following.

Firstly, the Cholesky decomposition has be applied on $\mathcal{H}(\pmb k)$ to find the complex matrix $\mathcal{K}$ that fulfills the equation
$\mathcal{H}(\pmb k) = \mathcal{K}^\dag \mathcal{K}$.
Then we could solve the eigenvalue problem of the Hermitian matrix $\mathcal{K}\Sigma\mathcal{K}^\dag$, and $\Lambda=\mathcal{V}^\dag \mathcal{K}\Sigma\mathcal{K}^\dag \mathcal{V}$, where $\mathcal{V}$ is an unitary matrix.
Finally, The bosonic Bogoliubov transformation matrix is given by
\beq\label{bBt}
\mathcal{U} = \mathcal{K}^{-1}\mathcal{V}(\Sigma \Lambda)^{1/2}.
\eeq
which diagonalizes spin-wave Hamiltonian $\mathcal{U}^\dag \mathcal{H}(\pmb k)\mathcal{U}=\Lambda$, $\Psi=\mathcal{U} \Phi$, where Bogoliubov quasi-particle basis $\Phi=(\alpha_{\pmb k}, \beta_{\pmb k}, \alpha_{\pmb {-k}}^\dag, \beta_{\pmb {-k}}^\dag)^\mathsf{T}$.
The form (\ref{bBt}) is much more useful to construct various effective actions in Appendix \ref{EFT}.

In the spin-wave spectrum, the momentum at a gapless point is carried by a specific in-plane magnetic order.
If $\delta\neq0$, $C_3$ symmetry is broken. The inequivalent $\pmb M$ points mean stripe orders and zigzag orders, respectively. Therefore we can distinguish the condensate momentum of AFM, stripe, and zigzag orders. As shown in Fig.\ref{fig:SW}, they condense at $\pmb \Gamma$, $\pmb M_1$, and $\pmb M_2$, respectively.

\section{ SU(4) Spin-Wave Spectrum at Low Fields}\label{app:lowfield}

In the low field region, the ground state for small $\Gamma'$ is a singlet state composed of the direct product of singlet dimers. Noticing that each dimer contains two spins $S=\frac{1}{2}$ which span a 4-dimensional Hilbert space, so we regard each dimer as an effective site placed with a SU(4) spin. In the following, we construct the SU(4) spin-wave theory on the square lattice formed by the effective sites.

Firstly, we introduce four species of  boson operators $$B^\dag= \big(b^{\dag}_{\up\up}, b^{\dag}_{\up\dn}, b^{\dag}_{\dn\up}, b^{\dag}_{\dn\dn}\big)$$ to represent the four bases $|\!\!\up\up\rangle,  |\!\!\up\dn\rangle, |\!\!\dn\up\rangle,  |\!\!\dn\dn\rangle$ on the dimer bond  respectively. Here we have omitted the site indices. This is the generalized Schwinger representation of  spins and the mapping is exact if the particle number constraint $B^\dag B=1$ is satisfied on each bond.

The spin operators on the A sublattice are represented by
\Beq
S^m_A = B^\dag {\sigma^m\over2}\otimes \sigma^0 B
\Eeq
with $\sigma^0$ the $2\times 2$ identity matrix, and the spin operators on the B sublattice are represented by
\Beq
S^m_B = B^\dag \sigma^0\otimes {\sigma^m\over2} B.
\Eeq

It is more convenient to combine these boson operators into the eigen modes $b_{S,m}$ with the total spin quantum number $S$ (for the two spins on the dimer bond) and its eigenvalue $m$ of the $S^z$-component, namely,
\Beq
\tilde B^\dag=\begin{pmatrix}b_{0,0}^\dag & b_{1,1}^\dag & b_{1,0}^\dag & b_{1,-1}^\dag\end{pmatrix} = \begin{pmatrix}b_{\up\up}^\dag & b_{\up\dn}^\dag & b_{\dn\up}^\dag & b_{\dn\dn}^\dag \end{pmatrix}  U
\Eeq
with
$U=\begin{pmatrix} 0 & 1 & 0 & 0 \\ \frac{\sqrt{2}}{2} & 0 & \frac{\sqrt{2}}{2} & 0 \\ -\frac{\sqrt{2}}{2} & 0 & \frac{\sqrt{2}}{2} & 0 \\ 0 & 0 & 0 & 1 \end{pmatrix}$. The advantage of adopting these new bases is that when the Heisenberg interactions on the strong  bonds are dominating, we can approximately regard the ground state as the product of singlets on the strong bonds,
\beq\label{eq:Dimer}
|{\rm Dimer}\rangle= \prod_i (b_{0,0}^\dag)_i |{\rm vac}\rangle,
\eeq
here $|{\rm  vac}\rangle$ stands for the vacuum state and $i$ labels the new site index.

In the new bases, the spin operators are rewritten as
\Beq
S^m_A = \tilde B^\dag U^\dag {\sigma^m\over2}\otimes \sigma^0 U\tilde B
\Eeq
for the A-sublattice
and
\Beq
S^m_B = \tilde B^\dag U^\dag\sigma^0 \otimes  {\sigma^m\over2} U\tilde B
\Eeq
for the B-sublattice.

Then  we can rewrite the spin-spin interactions in forms of the boson operators $\tilde B$. For instance, on a strong bond (namely the $z$-bond), the Heisenberg interaction $\pmb S^A\cdot \pmb S^B$ is represented as
\Beq
\pmb S^A\cdot \pmb S^B  = - \frac{3}{4}b^{\dag}_{0,0} b_{0,0} + \frac{1}{4} \sum_m b^{\dag}_{1,m} b_{1,m}
\Eeq

Secondly, we will adopt the HP approximation\cite{Muniz_2014} using dimerized singlets (\ref{eq:Dimer}) as the ground state, namely,
$$b^{\dag}_{0,0} b_{0,0} = \sqrt{1-\sum\limits^{3}_{m=1}b^{\dag}_{1,m}b_{1,m}}   \approx 1.$$
Only keeping the quadratic term according to the boson operators, we obtain a quadratic Hamiltonian for the bosons.

Thirdly, performinng Fourier transformation, $b_{\boldsymbol{k} ,s,m}^{\dagger}=\frac{1}{\sqrt{L}} \sum_{\boldsymbol{r}} b_{\boldsymbol{r} ,s,m}^{\dagger} \exp [i \boldsymbol{k} \cdot \boldsymbol{r}]$, we could obtain quadratic form Hamiltonian in momentum space,
\beq\label{eq:HSWSU4}
H_{\rm SW} = \sum_{\pmb k} \Psi^\dagger(\pmb k) \mathcal{H}(\pmb k) \Psi(\pmb k),
\eeq
with $\Psi^\dagger(\pmb k)=\Big (b_{1,1}^{\dag}(\pmb k), b_{1,0}^{\dag}(\pmb k), b_{1,-1}^{\dag}(\pmb k), b_{1,1}(-\pmb k), b_{1,0}(-\pmb k),$  $b_{1,-1}(-\pmb k)\Big)$, and $\mathcal{H}(\pmb k)$ is the 6 by 6 matrix.

Finally, we could obtain magnon dispersions in Fig.\ref{fig:SU4} by diagonalizing the non-Hermitian matrix $\Sigma\mathcal{H}(\pmb k)$, where $\Sigma={\rm diag}(1,1,1,-1,-1,-1)$.
We also obtain the bosonic Bogoliubov transformation based on the form (\ref{bBt}) in the previous section.

\section{Effective Field Theory of Continuous Phase Transitions at Higher Critical Fields}\label{EFT}

As we discuss in the main text, the symmetry of the rotated spin Hamiltonian (\ref{RJGM}) is generated by
\begin{enumerate}
    \item Translation by one lattice site in $\pmb a_1$ or $\pmb a_2$ direction:
        $\rm{T}_{\pmb a_1}: S_i\to S_{i+\pmb a_1}$
        and $\rm{T}_{\pmb a_2}: S_i\to S_{i+\pmb a_2}$.
    \item Inversion symmetry:
        $\mathcal{P}: S_{i}\to S_{i'}$,
        where $i=(a_1,a_2, \alpha)$ and $i'=(-a_1,-a_2, \overline\alpha)$. Note that $\alpha$ denotes one sublattice and $\overline\alpha$ with the other sublattice.
    \item Spin-orbital rotation symmetry:
        $\mathcal{C}_2\mathcal{T}: S_{i}\to (-1)\times \mathcal{R}_{a}(\pi)S_{i''}$,
        where $i=(a_1,a_2)$ and $i''=(a_2,a_1)$.
\end{enumerate}
where $\mathcal{R}_{a}(\pi)$ is the 180$^\circ$ rotation along $a$ direction (or $z$-bond) and $\mathcal{T}$ is time reversal symmetry.

Some of these symmetries are broken in the canted intermediate phases such as AFM$_{ab}$, stripe, and zigzag, but preserved in the high-field polarized phase (PP) and the low-field dimerized state. They are quite crucial to construct the corresponding effective actions to be presented in the following.

\subsection{Phase transition from PP to AFM$_{ab}$ order}

If the $\alpha_\mathbf{k}$ magnon condensation which leads to the QPT from PP to the canted AFM$_{ab}$ at $\pmb B_{critical}$, the corresponding order parameter takes the form:
\begin{equation}
\langle \alpha_\mathbf{k}\rangle =\psi\delta_{\mathbf{k},\mathbf{\Gamma}},\quad
\langle \beta_\mathbf{k}\rangle =0.
\end{equation}
where $\mathbf{\Gamma}=\pmb b_1 + \pmb b_2$ (namely, $\pmb \Gamma$ point in second BZ) and $\psi$ is the complex order parameter.

By introducing the bosonic Bogoliubov transformation (\ref{bBt}) in LSWT as
\begin{align}
    a_\mathbf{k}=u(\mathbf{k})\alpha_\mathbf{k}
    +v(\mathbf{k})\alpha_{-\mathbf{k}}^\dagger
    +u'(\mathbf{k})\beta_{\mathbf{k}}
    +v'(\mathbf{k})\beta_{-\mathbf{k}}^\dagger.
\end{align}
we establish the connection between the in-honeycomb-plane transverse quantum spin and the one complex order parameter:
\begin{align}
    \langle S_i^+\rangle
    \propto
    \langle a_i\rangle
    \propto
    u(\mathbf{\Gamma})\psi e^{i \mathbf{\Gamma} \cdot R_i} + v(\mathbf{\Gamma})\psi^* e^{-i \mathbf{\Gamma} \cdot R_i}.
\label{pmk0y}
\end{align}
Due to $u(\mathbf{\Gamma})= 0, v(\mathbf{\Gamma}) \neq 0$ in the form (\ref{bBt}) at the critical point, the order parameter is complex.

Because PP breaks no symmetry of the Hamiltonian, so one can study how the complex order parameter $\psi$ transform under the symmetries of the Hamiltonian:
\begin{enumerate}
    \item Translation symmetry:
        $\rm{T}_{a_1}: \psi(a_1,a_2)\to \psi(a_1,a_2)$
        and $\rm{T}_{a_2}: \psi(a_1,a_2)\to \psi(a_1,a_2)$;
    \item Inversion symmetry:
        $\mathcal{P}:
        \psi(a_1,a_2)
        \to \psi(-a_1,-a_2)$;
    \item Spin-orbital rotation symmetry:
        $\mathcal{C}_2\mathcal{T}:
        \psi(a_1,a_2)
        \to -i\psi^*(a_2,a_1)$.
\end{enumerate}
The transformation of the order parameter $\psi$ under spin-orbital rotation symmetry ($\mathcal{C}_2\mathcal{T}$) is a little subtle due to anti-unitary time-reversal symmetry ($\mathcal{T}$).
Notably, the time-reversal symmetry operates before and after the operator $\hat O$ in the quantum state $|\Psi\rangle$  to satisfy the following relations:
\beq
\langle  \widetilde \Psi | \mathcal{T} \hat O \mathcal{T}^{-1}| \widetilde \Psi \rangle = \langle   \Psi | \hat O^\dag |  \Psi \rangle, \nonumber
\eeq
where $|\widetilde \Psi \rangle = \mathcal{T} |\Psi \rangle$.
For simplicity, we define $\Lambda= \langle \Psi | S^+ | \Psi \rangle$, where $S^+=S^x + i S^y$. Thus we obtain
\beq
\mathcal{T} \Lambda &=& \langle \widetilde \Psi | S^+ | \widetilde \Psi \rangle = \langle \Psi | [\mathcal{T} S^+ \mathcal{T}^{-1} ]^\dag |\Psi \rangle =  \langle (\mathcal{T} S^+ \mathcal{T}^{-1})  \Psi | \Psi \rangle \nonumber \\
&=& - \langle S^{-} \Psi | \Psi \rangle  = - \langle \Psi | S^+ | \Psi \rangle = -\Lambda. \nonumber
\eeq
Because $\mathcal{C}_2$ is the 180$^\circ$ rotation along $a$ direction (or [1$\overline{1}$0] direction),  the transformation of operator $S^+$ ($S^-$) is $\mathcal{C}_2 S^+ \mathcal{C}_2^{-1} = -i S^{-}$ ($\mathcal{C}_2 S^- \mathcal{C}_2^{-1} = i S^+$).
In the end, we obtain the relation
\beq
\mathcal{C}_2\mathcal{T} \Lambda &=& \langle \widetilde \Psi | \mathcal{C}_2 S^+ \mathcal{C}_2^{-1} | \widetilde \Psi \rangle  = \langle \Psi | [\mathcal{C}_2 \mathcal{T}  S^+ \mathcal{T}^{-1} \mathcal{C}_2^{-1}  ]^\dag |\Psi \rangle \nonumber \\
&=&  \langle (\mathcal{C}_2\mathcal{T}  S^+ \mathcal{T}^{-1} \mathcal{C}_2^{-1} )  \Psi | \Psi \rangle = -i \langle \Psi | S^{-}  | \Psi \rangle = -i \Lambda^*. \nonumber
\eeq
Thus, in general cases, the spin-orbital rotation symmetry is $\mathcal{C}_2\mathcal{T}: \psi(a_1,a_2) \to -i\psi^*(a_2,a_1)$.

Due to the above symmetry analysis and considering linear spin-wave calculations, the following effective action with the dynamic exponent $z=2$ in the continuum limit is
\begin{align}
    \mathcal{S}
    =\int d\tau d^2r
    &\Big(\psi^*\partial_\tau\psi
    +v_{a_1}^2|\partial_{a_1}\psi|^2
    +v_{a_2}^2|\partial_{a_2}\psi|^2 \nonumber \\
    &-\mu |\psi|^2+U |\psi|^4 \Big).
\end{align}
Our microscopic calculation shows that $ \mu=B_{critical}-B_c$ and $U>0$.

The action at the fixed point is invariant under the scale transformation
\begin{align}\label{scaling}
    r'=re^{-l},\quad
    \tau'=\tau e^{-zl}, \quad
    \psi'=\psi  e^l.
\end{align}
Therefore, the scaling dimension of the parameter $U$ is zero (namely, the marginal operator).
If $\xi$ is the magnetic correlation length that diverges at the QCP, the following scaling relations hold
\begin{align}\label{dynamic}
    \mu \propto \frac{1}{\xi^{1/\nu}},\quad
    \Delta \propto \frac{1}{\xi^z}.
\end{align}
Due to $\Delta \propto |\mu|$, we could obtain the value of correlation length critical exponent $\nu=1/2$ .
And the universality class for the QPT is the same as the $z=2$ 2D Superfluid-Mott insulator transition.

At the mean-field level, we can substitute $\psi\to\sqrt{\rho_0}e^{i\phi_0}$ into the effective action
\begin{align}
    \mathcal{S}=-\mu\rho_0+U\rho_0^2
\end{align}
When $\mu= B_{critical}-B_c <0$, it is in the PP with $ \langle \psi \rangle =0$.
When $\mu>0$, it is in the canted AFM$_{ab}$ phase with  $ \langle \psi \rangle=\sqrt{\rho_0}e^{i\phi_0}$.

When $\mu<0$, $\langle \psi \rangle=0$ in the PP expanding the action upto second order:
\begin{align}
    \mathcal{S}
    =\int d\tau d^2r
    \Big(\psi^*\partial_\tau\psi
    +v_{a_1}^2|\partial_{a_1}\psi|^2
    +v_{a_2}^2|\partial_{a_2}\psi|^2
    -\mu |\psi|^2 \Big).
\end{align}
which lead to the gapped mode
\begin{align}
    \omega_k= -\mu+v_{a_1}^2k_{a_1}^2+v_{a_2}^2k_{a_2}^2
\label{gapleft}
\end{align}

In the canted AFM$_{ab}$ phase, $\mu>0$, we can write the fluctuations in the polar coordinates
$\psi=\sqrt{\rho_0+\delta\rho}e^{i(\phi_0+\delta\phi)}$
and expand the action up to the second order in the fluctuations:
\begin{eqnarray}
    \mathcal{S}
     &=\int d\tau d^2r
    \Big( i\delta\rho \partial_\tau\delta\phi
    +\frac{1}{4\rho_0}[v_{a_1}^2(\partial_{a_1}\delta\rho)^2+v_{a_2}^2(\partial_{a_2}\delta\rho)^2] \nonumber \\
    &+U(\delta\rho)^2
    +\rho_0[v_{a_1}^2(\partial_{a_1}\delta\phi)^2
         +v_{a_2}^2(\partial_{a_2}\delta\phi)^2]
    \Big).
\end{eqnarray}
Integrating out $\delta\rho$ leads to
\begin{align}
    \mathcal{S}
    =\int d\tau d^2r
    &\Big(
    \frac{1}{4U}(\partial_\tau\delta\phi)^2
    +\rho_0[v_{a_1}^2(\partial_{a_1}\delta\phi)^2 \nonumber \\
    &+v_{a_2}^2(\partial_{a_2}\delta\phi)^2]
    \Big).
\end{align}
It leads to the exotic Goldstone mode due to the emergent U(1) symmetry (in the classical degenerate ground state manifold although $\Gamma'\neq 0$) breaking:
\begin{align}
    \omega_k
    =\sqrt{4U\rho_0(v_{a_1}^2k_{a_1}^2+v_{a_2}^2k_{a_2}^2)}
\label{Goldhc1}
\end{align}

\subsection{Phase transition from PP to zigzag order}

If the $\alpha_\mathbf{k}$ magnon condensation which leads to the QPT from PP to the canted zigzag at $\pmb B_{critical}$, the corresponding order parameter takes the form:
\begin{equation}
\langle \alpha_\mathbf{k}\rangle =\tilde{\psi}\delta_{\mathbf{k},\mathbf{M}_2},\quad
\langle \beta_\mathbf{k}\rangle =0.
\end{equation}
where $\mathbf{M}_2=\frac{1}{2} \pmb b_1 + \frac{1}{2} \pmb b_2$ and $\tilde{\psi}$ is the complex order parameter.

One must use the Bogoliubov transformation to establish the connection between the transverse quantum spin and the one complex order parameter:
\begin{align}
    \langle S_i^+\rangle
    \propto
    \langle a_i\rangle
    \propto
    u(\mathbf{M}_2)\tilde{\psi} e^{i \mathbf{M}_2 \cdot R_i} + v(\mathbf{M}_2)\tilde{\psi}^* e^{-i \mathbf{M}_2 \cdot R_i}
\label{pmk0y}
\end{align}
Due to $u(\mathbf{M}_2)=v(\mathbf{M}_2)$ in the form (\ref{bBt}) at the critical point, we could redefine the order parameter $\langle a_i\rangle \propto \psi e^{i \mathbf{M}_2 \cdot R_i}$, where $\psi=\tilde{\psi}+\tilde{\psi}^*$ is real.

Because PP breaks no symmetry of the Hamiltonian, so one can study how the real order parameter $\psi$ transform under the symmetries of the Hamiltonian:
\begin{enumerate}
    \item Translation symmetry:
        $\rm{T}_{a_1}: \psi(a_1,a_2)\to -\psi(a_1,a_2)$
        and $\rm{T}_{a_2}: \psi(a_1,a_2)\to -\psi(a_1,a_2)$;
    \item Inversion symmetry:
        $\mathcal{P}:
        \psi(a_1,a_2)
        \to \psi(-a_1,-a_2)$;
    \item Spin-orbital rotation symmetry:
        $\mathcal{C}_2\mathcal{T}:
        \psi(a_1,a_2)
        \to -i\psi(a_2,a_1)$.
\end{enumerate}

Due to the above symmetry analysis and considering linear spin-wave calculations, the following effective action with the dynamic exponent $z=1$ in the continuum limit is
\begin{align}
    \mathcal{S}
    =\int d\tau d^2r
    &\Big((\partial_\tau\psi)^2
    +v_{a_1}^2(\partial_{a_1}\psi)^2
    +v_{a_2}^2(\partial_{a_2}\psi)^2 \nonumber \\
    &-\mu \psi^2
    +U \psi^4 \Big).
\end{align}
Our microscopic calculation shows that $\mu=B_{critical}-B_c$ and $U>0$.

The action at the fixed point is invariant under the scale transformation (\ref{scaling}).
Due to $\Delta \propto \sqrt{|\mu|}$, we could obtain the value of correlation length critical exponent $\nu=1/2$ through the form (\ref{dynamic}).
And the universality class for the QPT may be the $z=1$ 3D Ising universality class.

At the mean-field level, we can substitute $\psi\to\sqrt{\rho_0}$ into the effective action
\begin{align}
    \mathcal{S}=-\mu\rho_0+U\rho_0^2
\end{align}
When $\mu= B_{critical}-B_c <0$, it is in the PP with $ \langle \psi \rangle =0$.
When $\mu>0$, it is in the canted zigzag phase with  $ \langle \psi \rangle=\sqrt{\rho_0}$.

When $\mu<0$, $\langle \psi \rangle=0$ in the PP expanding the action upto second order:
\begin{align}
    \mathcal{S}
    =\int d\tau d^2r
    &\Big((\partial_\tau\psi)^2
    +v_{a_1}^2(\partial_{a_1}\psi)^2
    +v_{a_2}^2(\partial_{a_2}\psi)^2 \nonumber \\
    &-\mu \psi^2 \Big).
\end{align}
which lead to the gapped mode
\begin{align}
    \omega_k=\sqrt{-\mu+v_{a_1}^2k_{a_1}^2+v_{a_2}^2k_{a_2}^2}
\label{gapleft}
\end{align}

In the canted zigzag phase, $\mu>0$, we can write the fluctuations $\psi=\sqrt{\rho_0+\delta\rho}$
and expand the action up to the second order in the fluctuations:
\begin{eqnarray}
    \mathcal{S}
     =  \frac{1}{2\rho_0}\int d\tau d^2r
    &\Big( (\partial_\tau\delta\rho)^2
    +[v_{a_1}^2(\partial_{a_1}\delta\rho)^2+v_{a_2}^2(\partial_{a_2}\delta\rho)^2] \nonumber \\
    &+4\rho_0U(\delta\rho)^2    \Big).
\label{GoldHiggsact}
\end{eqnarray}
 which leads to one gapped Higgs mode
\begin{eqnarray}
    \omega_k
     =  \sqrt{4\rho_0U+v_{a_1}^2k_{a_1}^2+v_{a_2}^2k_{a_2}^2}
\end{eqnarray}

\subsection{Phase transition from PP to stripe order}

As shown in Fig.\ref{fig:SW}(b), there are two inequivalent momentum points at the higher critical field, namely $\pmb M_1$ and $\pmb M_3=\mathcal{R}_a(\pi)\pmb M_1$.
However, in the single-$\pmb Q$ framework, we should break the symmetry of two inequivalent momentum points by some small perturbations.
For example, the corresponding order parameter takes the form (if $\pmb M_3$ is gapped at the critical point):
\begin{equation}
\langle \alpha_\mathbf{k}\rangle =\tilde{\psi}\delta_{\mathbf{k},\mathbf{M}_1},\quad
\langle \beta_\mathbf{k}\rangle =0.
\end{equation}
where $\mathbf{M}_1=\frac{1}{2} \pmb b_1$ and $\tilde{\psi}$ is the complex order parameter.

One must use the Bogoliubov transformation to establish the connection between the transverse quantum spin and the one complex order parameter:
\begin{align}
    \langle S_i^+\rangle
    \propto
    \langle a_i\rangle
    \propto
    u(\mathbf{M}_1)\tilde{\psi} e^{i \mathbf{M}_1 \cdot R_i} + v(\mathbf{M}_1)\tilde{\psi}^* e^{-i \mathbf{M}_1 \cdot R_i}
\label{pmk0y}
\end{align}
Due to $u(\mathbf{M}_1)=v(\mathbf{M}_1)$ in the form (\ref{bBt}) at the critical point, we could redefine the order parameter $\langle a_i\rangle \propto  u(\mathbf{M}_1)\psi e^{i \mathbf{M}_1 \cdot R_i}$, where $\psi=\tilde{\psi}+\tilde{\psi}^*$ is real.

Because PP breaks no symmetry of the Hamiltonian, so one can study how the real order parameter $\psi$ transform under the symmetries of the Hamiltonian:
\begin{enumerate}
    \item Translation symmetry:
        $\rm{T}_{a_1}: \psi(a_1,a_2)\to -\psi(a_1,a_2)$
        and $\rm{T}_{a_2}: \psi(a_1,a_2)\to \psi(a_1,a_2)$;
    \item Inversion symmetry:
        $\mathcal{P}:
        \psi(a_1,a_2)
        \to \psi(-a_1,-a_2)$;
    \item Spin-orbital rotation symmetry:
        $\mathcal{C}_2\mathcal{T}:
        \psi(a_1,a_2)
        \to -i\psi(a_2,a_1)$.
\end{enumerate}

Due to the above symmetry analysis and considering linear spin-wave calculations, the following effective action with the dynamic exponent $z=1$ in the continuum limit is
\begin{align}
    \mathcal{S}
    =\int d\tau d^2r
    &\Big((\partial_\tau\psi)^2
    +v_{a_1}^2(\partial_{a_1}\psi)^2
    +v_{a_2}^2(\partial_{a_2}\psi)^2 \nonumber \\
    &-\mu \psi^2
    +U \psi^4 \Big).
\end{align}
Our microscopic calculation shows that $\mu=B_{critical}-B_c$ and $U>0$.
The universality class for the QPT from PP to stripe phase is the same as the QPT from PP to zigzag phase above.


At the mean-field level, we can substitute $\psi\to\sqrt{\rho_0}$ into the effective action
\begin{align}
    \mathcal{S}=-\mu\rho_0+U\rho_0^2
\end{align}
When $\mu= B_{critical}-B_c <0$, it is in PP with $ \langle \psi \rangle =0$.
When $\mu>0$, it is in the canted stripe phase with  $ \langle \psi \rangle=\sqrt{\rho_0}$.

When $\mu<0$, $\langle \psi \rangle=0$ in PP expanding the action upto second order:
\begin{align}
    \mathcal{S}
    =\int d\tau d^2r
    &\Big((\partial_\tau\psi)^2
    +v_{a_1}^2(\partial_{a_1}\psi)^2
    +v_{a_2}^2(\partial_{a_2}\psi)^2 \nonumber \\
    &-\mu \psi^2 \Big).
\end{align}
which lead to the gapped mode
\begin{align}
    \omega_k=\sqrt{-\mu+v_{a_1}^2k_{a_1}^2+v_{a_2}^2k_{a_2}^2}
\label{gapleft}
\end{align}

In the canted stripe phase, $\mu>0$, we can write the fluctuations $\psi=\sqrt{\rho_0+\delta\rho}$
and expand the action up to the second order in the fluctuations:
\begin{eqnarray}
    \mathcal{S}
     =  \frac{1}{2\rho_0}\int d\tau d^2r
     &\Big( (\partial_\tau\delta\rho)^2
    +[v_{a_1}^2(\partial_{a_1}\delta\rho)^2+v_{a_2}^2(\partial_{a_2}\delta\rho)^2] \nonumber \\
    &+4\rho_0U(\delta\rho)^2    \Big).
\label{GoldHiggsact}
\end{eqnarray}
 which leads to one gapped Higgs mode
\begin{eqnarray}
    \omega_k
     =  \sqrt{4\rho_0U+v_{a_1}^2k_{a_1}^2+v_{a_2}^2k_{a_2}^2}
\end{eqnarray}

Due to the same symmetry of the high-field PP and the low-field QDM, the effective field theory of the phase transition from QDM to intermediate ordered phases at lower critical fields is same as the phase transition from the high-field PP to intermediate ordered phases as we discussed above.
Furthermore, we could obtain corresponding effective actions describing other continuous phase transitions in the global phase diagram
by the similar approach.

\bibliography{BEC}

\end{document}